\documentclass[twocolumn]{aastex631}

\newcommand{\Var}{\ensuremath{\mathbf{Var}\{}}
\newcommand{\beq}{\begin{equation}}
\newcommand{\eeq}{\end{equation}}

\newcommand{\dpss}[2]{\,w_{#2}^{(#1)}}

\usepackage{amsmath}
\usepackage{graphicx}
\usepackage{amssymb}
\usepackage{natbib}

\newcounter{alignfirst}

\graphicspath{{./}{figures/}}

\begin{document}

\title[Optimal frequency-domain analysis for spacecraft time series]{Optimal frequency-domain analysis for spacecraft time series: Introducing the missing-data multitaper power spectrum estimator}






\author[0000-0002-8796-4974]{Sarah Dodson-Robinson}
 \affiliation{University of Delaware \\
 Bartol Research Institute \\
 217 Sharp Lab \\
 Newark, DE 19716, USA}

\author[0000-0003-3996-773X]{Charlotte Haley}
 \affiliation{Argonne National Laboratory \\
 Mathematics and Computer Science Division \\
 9700 S Cass Ave \\
 Lemont, IL 60439, USA}



\begin{abstract}

While the Lomb-Scargle periodogram is foundational to astronomy, it has a significant shortcoming: {\it the variance in the estimated power spectrum} does not decrease as more data are acquired. Statisticians have a 60-year history of developing variance-suppressing power spectrum estimators, but most are not used in astronomy because they are formulated for time series with uniform observing cadence and without seasonal or daily gaps. Here we demonstrate how to apply the missing-data multitaper power spectrum estimator to spacecraft data with uniform time intervals between observations but missing data during thruster fires or momentum dumps. The F-test for harmonic components may be applied to multitaper power spectrum estimates to identify statistically significant oscillations that would not rise above a white noise-based false alarm probability. Multitapering improves the dynamic range of the power spectrum estimate and suppresses spectral window artifacts. 
We show that the multitaper--F-test combination applied to {\it Kepler} observations of KIC~6102338 detects differential rotation without requiring iterative sinusoid fitting and subtraction. Significant signals reside at harmonics of both fundamental rotation frequencies and suggest an antisolar rotation profile. {Next we use the missing-data multitaper power spectrum estimator to identify the oscillation modes responsible for the complex ``scallop shell'' shape of the K2 light curve of EPIC~203354381.}
We argue that multitaper power spectrum estimators should be used for all time series with regular observing cadence.

\end{abstract}



\section{Introduction}
\label{sec:intro}




Starting with \citet{schuster1898, schuster06}, astronomers investigating periodic phenomena have relied on statistical methods for estimating power spectra.\footnote{For a formal definition the power spectral representation of a stationary random process, see \citet{cramer40, yaglom62, pw2020}.} {The standard Schuster periodogram is
\begin{equation}
    \hat{S}^p(f) = {\frac{1}{N}} \left| \sum_{t=0}^{N-1} x_n e^{-2 \pi i f {t_n}} \right|^2,
    \label{eq:rawperiodogram}
\end{equation}
where $x_n$ is the observed time series, $n$ is the time index, $N$ is the number of observations, $f$ is frequency, $t_n$ is time, and $\hat{\cdot}$ denotes a statistical estimator.} Although there exists no unbiased, nonparametric power spectrum estimator $\hat{S}(f)$, {the Schuster periodogram is {\it asymptotically unbiased}, meaning it approaches the true physical power spectrum $S(f)$ as the number of observations approaches infinity \citep{pw2020}:
\begin{align}
\operatorname{Bias}\{ \hat{S}(f) \} &= E \{\hat{S}(f) \} - S(f) \\
\lim_{N\rightarrow \infty}  \operatorname{Bias} \{ \hat{S}^p(f) \} &= 0 
\label{eq:biasdef}
\end{align}
(where $E \{ \}$ is expected value).} But the number of observations required for the Schuster periodogram to yield an acceptably accurate power spectrum estimate is often too large to be practical: \citet{th14} show a barometric pressure time series with $N = 1000$ for which the Schuster periodogram is biased at high frequencies by seven orders of magnitude.

Ideally our power spectrum estimator would be consistent, such that its variance decreases with increasing sample size. Unfortunately, the variance of the Schuster periodogram follows
\begin{equation}
    \Var{\hat{S}^{p}(f)} \} = E \{|\hat{S}^p(f) - S(f)|^2 \} \propto S^2(f)
    \label{eq:varper}
\end{equation}
\citep[Ch.\ 9]{bloomfield00}. Since the right-hand side of Equation \ref{eq:varper} does not depend on $N$, the Schuster periodogram is {\it inconsistent}: its variance does not decrease as more observations are obtained. Any periodogram computed from a finite-length time series is therefore susceptible to variance-induced false positives. 

In the mid-to-late twentieth century, statisticians developed consistent, low-bias power spectrum estimators \citep{bartlett48, blackman1958measurement, welch67, thomson1982spectrum}. But the fact that ground-based observations always have uneven observing cadence has kept such power spectrum estimators out of astronomy, since they are formulated for time series with uniform observing cadence (constant $t_n - t_{n-1}$). Instead, astronomers typically use the Lomb-Scargle periodogram developed by \citet{lomb76} and \citet{scargle1982studies} and generalized by \citet{zechmeister09}. The Lomb-Scargle periodogram has facilitated decades of groundbreaking discoveries, including double degenerate binaries \citep{bragaglia90, marsh95}, sunlike magnetic activity cycles in nearby stars \citep{baliunas95}, and planets orbiting Proxima Centauri \citep{angladaescude16, damasso20, faria22}. {But the Lomb-Scargle periodogram shares the inconsistency of the Schuster periodogram and also inherits bias from the uneven observing cadence \citep{dawson10, vanderplas18}.} We seek an astronomical power spectrum estimator that is both minimally biased and consistent.




The gold-standard method for power spectral analysis of time series with uniform observing cadence, the multitaper \citep[][Ch.\ 8]{thomson1982spectrum, bronez92, stoicasundin99, pw2020}, {is a nonparametric, Frequentist method that optimally solves the bias and variance problems inherent to periodograms.} Multitaper suppresses variance by averaging together multiple independent, low-bias estimators of the power spectrum. Each independent power spectrum estimator $S^{(k)}_{xx}(f)$ comes from premultiplying $x_n$ with a set of weights or {\it taper} $w^{(k)}_n$ before computing the Fourier transform. With $K$ tapers, the variance of the final power spectrum estimator $\hat{S}_{xx}(f)$ is {lower than the periodogram variance (Equation \ref{eq:varper})} by a factor of almost $1/K$. For example, using $K = 8$ and a time-bandwidth product (to be defined in \S \ref{subsec:whatistaper}) of 5, one achieves a fivefold reduction of false positive peaks at all levels of statistical significance \citep{th14}.

The classical multitaper method of \citet{thomson1982spectrum} can only be applied to time series with uniform timesteps, a condition that is almost never true in astronomy {\citep[though multitaper has been used to investigate the {\it angular} power spectrum of cosmic microwave background fluctuations;][]{fowler10}}. However, \citet{bronez1988spectral} and \citet{chave2019multitaper} extended the method to time series sampled at regular intervals that are missing some observations. The Bronez-Chave technique is well suited to time series from missions such as {\it Kepler} and K2, in which regularly sampled time series are gapped when the reaction wheels dump momentum \citep{garcia11, tesshandbook}, the thrusters fire unpredictably \citep{saunders19}, or the observations suffer argabrightenings that the data pipeline can't remove \citep{jenkins10, tesshandbook}.

{Here we explain basic multitaper principles and} demonstrate how to apply the missing-data multitaper power spectrum estimator to spacecraft time series.\footnote{\citet{springford20} presented an interpolation-based multitaper estimator for observing cadences with small departures from uniformity which we discuss briefly in \S \ref{subsec:gaps}.} We introduce the mathematics of the classical multitaper method in \S \ref{sec:multitaper}, then demonstrate its extensions---tapers for time series with missing data, confidence intervals, and the F-test for identifying statistically significant oscillations---in \S \ref{sec:extensions}.
In \S \ref{sec:kepler_diffrot}, we {showcase both the accuracy and the precision of the missing-data multitaper method by using it to diagnose antisolar rotation in} {\it Kepler} observations of the young sunlike star KIC~6102338. {In \S \ref{sec:scallop}, we identify the set of modes that produce one of the complex ``scallop-shell'' light curves discovered by \citet{stauffer17} and \citet{stauffer18}.}
We present our conclusions in \S \ref{sec:conclusions}. A software package that performs classical and missing-data multitaper calculations, \texttt{Multitaper.jl}, is discussed on page \pageref{sec:software}.

\section{Multitaper basics: The classical multitaper method} 
\label{sec:multitaper}

The multitaper power spectrum estimator mitigates what \citet{percival94} described as the ``tragedy of the periodogram''.\footnote{Statistician John Tukey famously said, ``More lives are lost by looking at the raw periodogram than by any other action involving time series.''} As discussed in \S \ref{sec:intro}, the tragedy is twofold: periodograms are both biased and inconsistent. Here we introduce basic principles of multitaper power spectrum estimation. We begin with the physical motivation for tapering (\S \ref{subsec:whatistaper}), then proceed to taper calculation (\S \ref{subsec:eigencalc}). In \S \ref{subsec:sampletapers} we show a sample set of tapers, then in \S \ref{subsec:estimatespectrum} we describe how to estimate the power spectrum. We discuss bias and accuracy in \S \ref{subsec:accuracy} and provide guidance for bandwidth choice in \S \ref{subsec:bandwidth}. We finish the section with an example multitaper power spectrum estimate in \S \ref{subsec:classicalmpse}.



\setcounter{alignfirst}{\theequation}
\stepcounter{alignfirst}

\subsection{What is a taper?}
\label{subsec:whatistaper}

The classical multitaper method introduced by \cite{thomson1982spectrum} employs the discrete prolate spheroidal sequences (DPSS) discovered by \citet{s78} as tapers $w_n^{(k)}$ for $n = 0,\ldots,N-1$ and $k = 0,\ldots, K-1$. {Tapers are sequences that produce optimally shaped spectral windows for Fourier analysis. The expected value of the $k^{\rm th}$ tapered power spectrum estimator is
\begin{equation}
E \{ S^{(k)}_{xx}(f) \} = S(f) * W^{(k)}(f),
\label{eq:expectedspectrumestimate}
\end{equation} 
where 
$*$ denotes convolution and 
\begin{equation}
W^{(k)}(f) = \sum_{t=0}^{N-1} | w^{(k)}_n \exp [-2 \pi i f t] |^2
\label{eq:spectralwindow}
\end{equation}
is the $k^{\rm th}$ spectral window. The goal of tapering is to minimize biasing {\it leakage}, that is, spurious power at a given frequency that is actually attributable to power from a far-away frequency. The ideal spectral window is $W(f) = \delta(f)$, for which the expected value of the power spectrum estimator would equal the true power spectrum. However, there is no finite-length taper $w_n$ that Fourier transforms to $\delta(f)$---it's only possible to achieve $W(f) = \delta(f)$ for time series and tapers of infinite length. The boxcar function $w_n = \sqrt{N}$, which is the ``taper'' that results from having a finite-length dataset that is otherwise untapered, has a $\operatorname{sinc}^2$ function as a spectral window, which leaks power into its high sidelobes. For an introduction to tapering principles, see \citet{harris78}. See \citet[Chapter 8]{pw2020} and \citet[Chapter 8]{mudelsee14} for textbook descriptions of the DPSS tapers.

Since the DPSS are orthogonal, such that $w^{(k)}_n \cdot w^{(m)}_n = 0$ for all $k \neq m$, the $S^{(k)}_{xx}(f)$ are independent. Averaging together the $S^{(k)}_{xx}(f)$ therefore produces a {\it multitaper} power spectrum estimator with variance reduced by a factor of $\simeq 1/K$ from any single-taper power spectrum estimator.\footnote{Here is a simple analogy: the $1\sigma$ error on the estimated average $V$-band flux of a star is $\sigma_{\mu} = \mu / \sqrt{N}$, where $\mu$ is the sample mean of the $V$-band flux measurements and $N$ is the number of observations. Since $\sigma_{\mu}^2$ is the variance on the estimated mean, we find that $\operatorname{Var} \{ \mu \} = \sigma_{\mu}^2 = \mu^2 / N$: the variance is inversely proportional to the number of data points, given that each data point is an independent estimate of the $K$-band flux. With multitaper, each tapered power spectrum estimator plays the role of a single $V$-band observation.} We can quantify the quality of our power spectrum estimator using the {\it mean-squared error} (MSE):
\begin{equation}
\label{eq:MSE}
    \operatorname{MSE} = \Var{\hat{S(f)}} \} + \left[ \operatorname{Bias}\{ \hat{S}(f) \} \right]^2 
\end{equation}
\citep[Ch.\ 8]{pw2020}. Multitaper has the lowest MSE of any nonparametric power spectrum estimator, including the Schuster periodogram, the Lomb-Scargle periodogram, Bartlett's method \citep{bartlett48}, and Welch's method \citep{welch67, SDR2022}.
}

\subsection{Calculating the multitapers}
\label{subsec:eigencalc}


The DPSS tapers $w_n^{(k)}$ are defined by the eigenvalue equation
\begin{equation} 
    \sum_{t' = 0}^{N-1} \frac{\sin (2 \pi \varpi [t - t'])}{\pi [t - t']} w_{t'}^{(k)} = \lambda^{(k)} w_n^{(k)}.
    \label{eq:slepian}
\end{equation}
{To solve Equation \ref{eq:slepian}, all observation timestamps must be divided by the observing cadence $\Delta t_{\rm obs}$ in order to construct a time series with integer timestamps and unit sampling, $\Delta t = t_{n+1} - t_n = 1$.} When there are no missing values in the time series, the matrix $\sin (2 \pi \varpi [t - t']) / (\pi [t - t'])$ is Toeplitz and {commutes with a tridiagonal matrix, enabling the use of computationally efficient eigenvalue routines to solve for the eigenvectors $w^{(k)}_n$} \citep{grunbaum1981eigenvectors}. {In Equation \ref{eq:slepian}, $\varpi$ is the {\it bandwidth}, or half-width of the main lobe of the spectral window.\footnote{See \citet{harris78} for a detailed discussion of the bandwidths of commonly used tapers (also called windows), such as the Hamming taper, the Hann taper, and the Blackman-Harris taper.} Oscillations in the time series will manifest as peaks in the power spectrum estimate that have full width $2 \varpi$. We will discuss bandwidth selection in more detail in \S \ref{subsec:bandwidth}.} Throughout this paper we will use $\varpi$ to denote bandwidth in rescaled units where $\Delta t = 1$ and $\tilde{\varpi} = \varpi / \Delta t_{\rm obs}$ to refer to bandwidth in physical units.

The DPSS tapers have the desirable property of being maximally bandlimited, which means they provide the highest {\it spectral concentrations} inside $-\varpi < f < \varpi$ that can be obtained for any finite-length, discrete sequence. 
{In units where $\Delta t = 1$}, spectral concentrations are defined as
\begin{equation}
    \lambda^{(k)} = \frac{\int_{-\varpi}^{\varpi} W^{(k)}(f) \: df}{\int_{-1/2}^{1/2} W^{(k)}(f) \: df},
    \label{eq:spectralconcentration}
\end{equation}
where $1/2$ cycle per time unit is the maximum frequency probed by the time series (Nyquist frequency) and the $\lambda^{(k)}$ in Equation \ref{eq:spectralconcentration} are equal to the eigenvalues in Equation \ref{eq:slepian}. {Equation \ref{eq:spectralconcentration} measures the fraction of power in $W^{(k)}(f)$ that is confined to the frequency interval $-\varpi < f < \varpi$, so that $0 < \lambda^{(k)} < 1$. The higher the value of $\lambda^{(k)}$, the lower the bias contribution to the mean-squared error (Equation \ref{eq:MSE}; see also \S \ref{subsec:accuracy}).}

\subsection{A sample set of tapers}
\label{subsec:sampletapers}

\begin{figure*}
    \centering
    \includegraphics[width=0.7\textwidth]{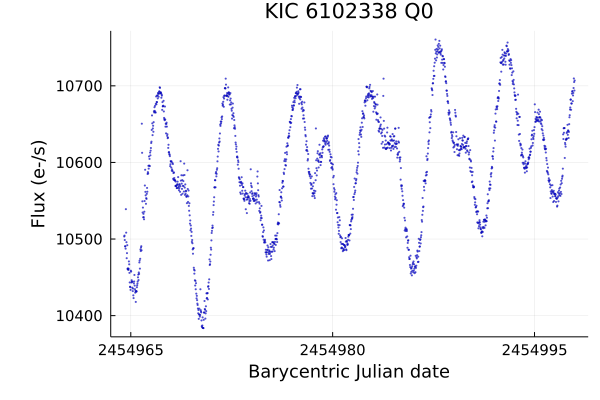}
    \caption{{\it Kepler} Q0 long-cadence observations of KIC~6102338 after replacement of missing data.}
    \label{fig:Kepler_Q0_fluxes}
\end{figure*}

Figure \ref{fig:Kepler_Q0_fluxes} shows the {\it Kepler} Q0 long-cadence observations of the differential rotator KIC~6102338. Out of $N = 1639$ flux measurements in the PDCSAP\footnote{PDCSAP = presearch data conditioning simple aperture photometry; see \citet{jenkins10}.} light curve downloaded from the MAST archive,\footnote{https://mast.stsci.edu/portal/Mashup/Clients/Mast/Portal.html} only 16 had been rejected by the {\it Kepler} pipeline; these were denoted NaN in the light curve table. {To construct a time series with uniform observing cadence and no missing data on which to demonstrate the classical multitaper technique, we used a simple interpolation technique where we replaced each NaN with the immediately preceding PDCSAP flux.} 

\begin{figure*}
    \centering
    \includegraphics[width=0.7\textwidth]{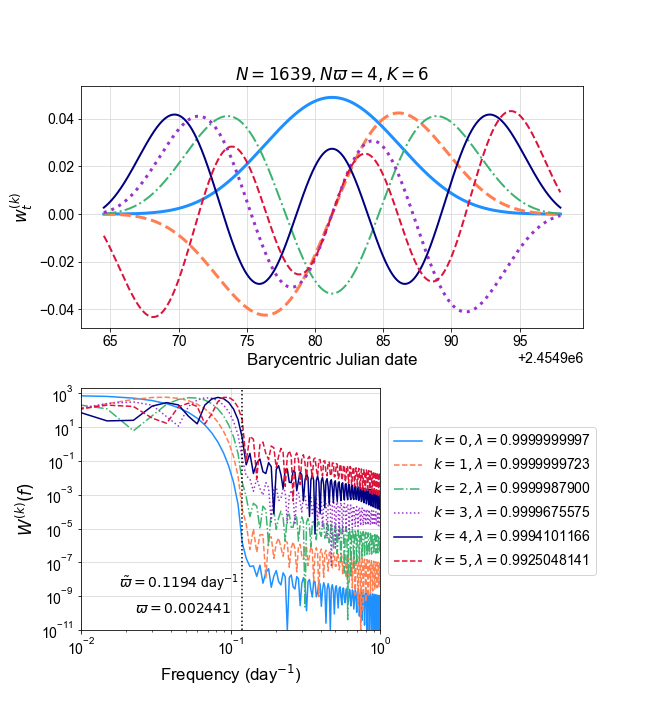}
    \caption{{Top:} The six DPSS tapers with the highest spectral concentrations given $N = 1639$, $N\varpi = 4$ ($\varpi = 0.002441$, $\tilde{\varpi} = 0.1194$~day$^{-1}$). {Bottom:} The positive half of the $K$ spectral windows $W^{(k)}(f)$ corresponding to each taper, with a vertical dotted line marking the bandwidth $\tilde{\varpi}$. The legend at the lower right shows the values of $k$ and $\lambda$ that correspond to each taper.}
    \label{fig:dpss}
\end{figure*}

Figure \ref{fig:dpss} depicts Equation \ref{eq:slepian} applied to the timestamps from the {\it Kepler} Q0 KIC~6102338 observations. The top panel of Figure \ref{fig:dpss} shows the six DPSS's $w^{(k)}_t$, $k = 0, \ldots, 5$ that yield the highest matrix eigenvalues $\lambda^{(k)}$. The bottom panel of Figure \ref{fig:dpss} shows the positive halves of the $K$ spectral windows $W^{(k)}(f)$ (spectral windows are symmetric about $f = 0$). Spectral concentrations are shown in the legend. The characteristics of the spectral window are very important, as we will see in \S \ref{subsec:windows}. {Note that the frequency axis in Figure \ref{fig:dpss} is logarithmic, so the main lobes of the spectral windows are quite narrow ($\varpi = 0.002441$, which covers only $0.48$\% of the frequency domain).}

\subsection{Estimating the power spectrum}
\label{subsec:estimatespectrum}

Recall that the orthogonal DPSS's of Equation \ref{eq:slepian} provide $K$ independent power spectrum estimators $\hat{S}^{(k)}_{xx}(f)$ (\S \ref{subsec:whatistaper}). Each $\hat{S}^{(k)}_{xx}(f)$ is a modified Schuster periodogram, or {\it eigenspectrum}, of $x_n$:
\begin{equation}
    \hat{S}^{(k)}_{xx}(f) = \left| \sum_{t=0}^{N-1} w^{(k)}_t x_n e^{-2 \pi i f t} \right|^2.
    \label{eq:taperedpowspec}
\end{equation}
The multitaper power spectrum estimate is a weighted average of the eigenspectra
\begin{equation} \label{eq:Eigenspec}
  \hat{S}_{xx}(f) = \frac{\sum_{k=0}^{K-1} d_k^2(f) \hat{S}_{xx}^{(k)}(f)}{\sum_{k=0}^{K-1} d_k^2(f)},
\end{equation}
where the averaging limits the variance of the estimator. At frequencies where the power spectrum is varying rapidly {(i.e.\ $|d \hat{S}_{xx}(f) / df|$ is high)}, the weighting adjusts to favor minimizing bias over variance by downweighting eigenspectra with lower spectral concentrations, {as leakage could otherwise render the estimated power spectrum slope artificially shallow. We will demonstrate how to calculate the bias of $\hat{S}_{xx}(f)$ in \S \ref{subsec:accuracy} and estimate the variance of $\hat{S}_{xx}(f)$ in \S \ref{subsec:jk}.} The frequency-dependent weights, $d(f)$, are chosen so that each $\hat{S}_{xx}^{(k)}(f)$ satisfies the equation
\begin{equation}\label{eq:weights}
  \sum_{k=0}^{K-1} \frac{\lambda_k[ \hat{S}_{xx}(f) - \hat{S}_{xx}^{(k)}(f)] }{
    (\lambda_k \hat{S}_{xx}(f) + s_x^2[1-\lambda_k])^2} = 0
\end{equation}
\citep[\S V]{thomson1982spectrum}, where $s_x^2$ is the sample variance of the time series:
\begin{equation}
    s_x^2 = \frac{1}{N-1} \sum_{t=0}^{N-1} x_n^2.
    \label{eq:samplevariance}
\end{equation}

{The coupled equations \ref{eq:Eigenspec} and \ref{eq:weights} are solved iteratively so as to produce estimates $\hat{S}_{xx}(f)$ that match within a user-defined tolerance. All functionality for calculating $\hat{S}_{xx}(f)$ according to Equations \ref{eq:Eigenspec} and \ref{eq:weights} is included in the \texttt{Multitaper.jl} software package \citep{multitaperpkg} used for this paper, as well as in the \texttt{R} multitaper package and the \texttt{python} package by \citet{prieto22}.} Multitaper outperforms the \citet{welch67} power spectrum estimator adapted for astronomical use by \citet{SDR2022}. If one holds two of bandwidth, bias, and variance at fixed values, the third quantity is smaller for the multitaper estimator than for any other power spectrum estimator \citep{bronez92}.

\subsection{Bias and accuracy}
\label{subsec:accuracy}

{Multitapering optimally minimizes the broadband bias defined in Equation \ref{eq:biasdef}. An upper limit to the broadband bias is}
\begin{equation}
    \operatorname{bias} \left \{ \hat{S}_{xx}(f) \right \} \leq \left( 1 - \lambda^{(K-1)} \right) s_x^2,
    \label{eq:mtbias}
\end{equation}
\citep{thomson1982spectrum}, where $\lambda^{(K-1)} = \operatorname{min}\{ \lambda^{(k)} \}$ is the spectral concentration of the highest-order DPSS and $\operatorname{min}\{ \cdot \}$ denotes minimum. In Figure \ref{fig:dpss}, our choice of $K = 6$ limits the bias to $0.0075 \, s_x^2$ ($\lambda^{(5)} = 0.9925$). Equation \ref{eq:mtbias} guarantees that the multitaper power spectrum estimate cannot be excessively far from the true power spectrum; no such guarantee exists for the Schuster and Lomb-Scargle periodograms. In practical application, multitaper estimators are likely to {have orders of magnitude lower MSE (Equation \ref{eq:MSE})} than any other nonparametric estimator when the dynamic range of the true power spectrum is large \citep[see example in][Figure 1]{th14}.

{Although astronomers often focus on estimating oscillation frequencies, there are applications for which accurately measuring the {\it power} in either the oscillations or the underlying broadband continuum is critical. For example, Alfv\'{e}n wave excitation in the solar photosphere is traced by the motion of magnetic ``bright points,'' which anchor flux tubes that reach up into the corona. From radiative magnetohydrodynamic simulations of granulation in the solar photosphere, \citet{vankooten17} predicted a bright-point motion power spectrum of $S(f) \propto f^{-1}$, in agreement with observations from the Swedish 1m Solar Telescope and the Solar Optical Telescope \citep{chitta12}. Without an accurate $E \{S(f) \}$ from observations, it would be much more difficult to assess the validity of the \citet{vankooten17} model at describing both the coronal heating rate and the granule size spectrum. Asteroseismic extraction of stellar parameters also relies on accurately representing both the total power and the shape of the Gaussian oscillation envelope \citep[e.g.][]{kallinger10}. In \S \ref{sec:kepler_diffrot} we show an example where accurately estimating the power in each detected oscillation is crucial to the science goal.}


\subsection{Why multitaper does not use error bars}
\label{subsec:errors}

\begin{figure*}
    \centering
    \includegraphics[width=0.95\textwidth]{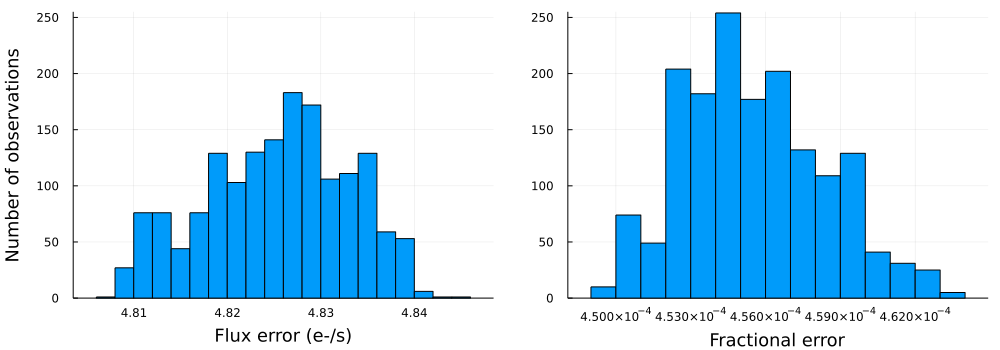}
    \caption{{Left:} Distribution of absolute flux errors in the {\it Kepler} Q0 observations of KIC~6102338. {Right:} Distribution of fractional flux errors in the same dataset.}
    \label{fig:Q0errors}
\end{figure*}

{It may seem unusual that Equations \ref{eq:taperedpowspec}--\ref{eq:weights} do not include uncertainties. Unfortunately, weighting the data based on nonuniform error bars, as in the generalized Lomb-Scargle periodogram and its Bayesian extension \citep{zechmeister09, mortier15}, produces a ``taper'' for which the spectral window is shaped by the error bars (Equation \ref{eq:spectralwindow}). The expected value of the power spectrum estimator then departs from true power spectrum in unquantifiable ways (Equation \ref{eq:expectedspectrumestimate}; see also the discussion of accuracy in \S \ref{subsec:accuracy}). Furthermore, the Bayesian and generalized Lomb-Scargle periodograms become dimensionless when computed with uncertainty-derived weights; instead of power density, the quantity estimated is either an approximate signal-to-noise ratio or a probability. Since the data are scaled by uncertainty-derived weights before the computation of trigonometric sums \citep{zechmeister09, mortier15}, there is no straightforward conversion back into physical units.  


If the observer's goal is to detect a small number of statistically significant oscillations, it is less critical that the power spectrum estimate be minimally biased across all frequencies. For example, radial velocity planet hunters will often model a time series as a sum of 1--5 Keplerian orbits plus a Gaussian process to treat stellar activity. For such applications, it is sufficient that the periodogram peak corresponding to each oscillation sticks out sufficiently far above the continuum; accurate, dimensional estimates of each signal's power are not required. (Note, however, that spectral leakage from strong oscillations can hide weaker oscillations, so some control over bias is important even for signal detection.) But as we learned in \S \ref{subsec:accuracy}, studying phenomena with broadband power spectra such as granulation and turbulence requires quantifying the contribution of sinusoids of {\it all} frequencies to the time series (not just the strongest oscillations).

Furthermore, we note that the error bars on most modern spacecraft photometry are both miniscule when compared to the quantity being measured and nearly uniform. The left panel of Figure \ref{fig:Q0errors} shows a histogram of $1\sigma$ errors on the {\it Kepler} Q0 observations of KIC~6102338 (Figure \ref{fig:KeplerQ0}), while the right panel contains a histogram of fractional errors from the same dataset. The distribution of absolute flux error spans $< 0.1$~e$^-$~s$^{-1}$, while the fractional flux error does not exceed $4.65 \times 10^{-4}$, in keeping with {\it Kepler}'s mandate to deliver photometry with parts-per-10,000 precision. {\it Twinkle}, a forthcoming Earth-orbiting satellite that will deliver high-cadence, low-resolution spectroscopy spanning 0.5--4.5$\micron$, will have $\sim 1$ part-per-thousand precision \citep{stotesbury22}. It is acceptable to leave out error bars when estimating the power spectra of such high-quality data, a practice followed by the {\it Kepler} Asteroseismology Program \citep[e.g.][]{kallinger10}. Finally, when the error bars are uniform, the weighted generalized Lomb-Scargle periodogram reduces to a standard Lomb-Scargle periodogram computed without uncertainty-based weights \citep{zechmeister09}.}

\subsection{Choice of bandwidth}
\label{subsec:bandwidth}

The choice of $N\varpi$ allows the observer to trade bias for variance. 
Equation \ref{eq:slepian} yields $2N\varpi$ maximally bandlimited {tapers $w_n^{(k)}$ with $\lambda \simeq 1$} \citep{thomson1982spectrum}. High-order DPSS with $k > 2N\varpi$ have $\lambda \ll 1$, so do not make useful tapers; {the maximum allowable value of $K$ is therefore $2N\varpi$}. To improve variance suppression, {the observer can choose a larger bandwidth so as to increase $K$}. To prioritize bias suppression, {which is important if the physical process under study produces a power spectrum with high dynamic range}, the observer should use a narrower bandwidth and fewer tapers. The control of the bias-variance tradeoff through the adjustment of bandwidth is widely recognized as the main performance advantage of the multitaper method \citep{bronez92,haley17}. For more details about how adjustment of the bandwidth can be used to trade bias for variance, see \citet{pw2020}.



To construct a multitaper power spectrum estimate, the observer sets $\varpi$ such that the time-bandwidth product $N\varpi$ is a small integer or half-integer, which creates a full bandwidth $2\varpi$ that is an integer number of Rayleigh resolution units. As in optics, the Rayleigh resolution $\mathcal{R}$ gives the minimum frequency separation required for two oscillations to be statistically distinguishable:
\begin{equation}
    \mathcal{R} = \frac{1}{t_{N-1} - t_0}
    \label{eq:Rayleigh}
\end{equation}
\citep{rayleigh1903}. 
In Figure \ref{fig:dpss}, $N\varpi = 4$---{a standard choice in the multitaper literature for time series with $N \sim 1000$ \citep[e.g.][]{thomson1982spectrum, ojeda02}}---and the main lobe of the spectral window is $2N\varpi = 8$ Rayleigh units wide. Next the observer chooses the optimum integer value of $K$ in the range $N\varpi \leq K \leq 2N\varpi$. For example, Figure \ref{fig:dpss} could include two more bandlimited DPSS's than are pictured ($k = 6, k = 7$), but observers are not required to use all DPSS's.
The spectral concentration of $w^{(7)}_t$ is $\lambda^{(7)} = 0.6988$, which means more than 30\% of the power in an oscillation with frequency $f_0$ leaks out of the interval $f_0-\varpi < f_0 < f_0+\varpi$. Since $\lambda^{(6)} = 0.9367$, {which allows more than 6\% spectral leakage outside of the bandwidth}, an especially bias-averse observer might set $K = 6, k = 0 \ldots 5$, as we have done. A geophysics example that prioritizes low bias is the measurement of the elastic thickness of the central Australian lithosphere, for which \citet{swain03} used $N\varpi = 2$, $K = 2$.

To maintain the consistency of the multitaper estimator, it is important to increase K (and accordingly $N\varpi$) along with the sample size. \citet{haley17} present a method for finding the bandwidth {that minimizes the MSE} using the second derivative of a spline fit to $\hat{S}_{xx}(f)$ on a frequency grid with spacing $2\varpi$. Heuristically, their experiments with a Lorentzian power spectrum yield optimal time-bandwidth products $N\varpi = 5$ for $N = 500$, $N\varpi = 8$ for $N = 1000$, and $N\varpi = 11$ for $N = 1500$. However, the best choice of $N\varpi$ depends on the process power spectrum and the science goal. {Since all applications in this manuscript involve detecting periodic signals,} we use smaller time-bandwidth products that allow more precise oscillation frequency estimation. {Seismologists often use narrow bandwidths (small $N\varpi$) in order to detect oscillations: \citet{chave2019ocean} selected $N\varpi = 8$, $K = 15$ to identify closely spaced seismic modes in a time series of $N = 597,504$ deep-ocean pressure measurements, while \citet{perez07} chose $N\varpi = 3$, $K = 5$ in their calculation of lithospheric elastic thickness from seismic tomography. For astronomical time series with $1000 \lessapprox N \lessapprox 5000$, we recommend values of $N\varpi$ between 4 and 8. For $N \lessapprox 1000$, $N\varpi = $3--4 is a good starting point. \citet{thomson1982spectrum} recommends that the observer use higher $N\varpi$ to characterize the continuum and lower $N\varpi$ to study the oscillations, so that a detailed multitaper analysis of a single time series may involve several power spectrum estimates with different $N\varpi$.}



\subsection{A classical multitaper power spectrum estimate}
\label{subsec:classicalmpse}

\begin{figure*}
    \centering
    \includegraphics[width=0.7\textwidth]{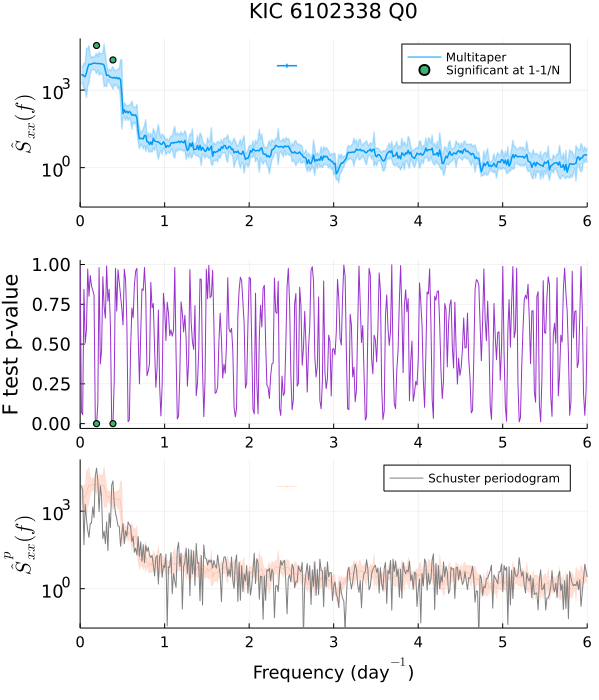}
    \caption{{Top:} Multitaper power spectrum estimate from the {\it Kepler} Q0 observations of KIC~6102338, using the tapers shown in Figure \ref{fig:dpss}. The light blue shading shows the 95\% confidence interval. The vertical crossbar shows the expected value of the jackknife variance estimate $\widehat{\Var} \hat{S}_{xx}(f) \}$, while the horizontal crossbar has width $\tilde{\varpi}$. Green dots indicate signals that exceed the $1-1/N$ significance threshold (\S \ref{subsec:Ftest}). {Middle:} The p-value given by the F-test for harmonic components, which is performed at each frequency (\S \ref{subsec:Ftest}). Green dots mark the frequencies for which $p < 1/N$.
    {Bottom:} The Schuster periodogram (black) overplotted on the multitaper 95\% confidence intervals (orange). The Schuster periodogram has larger variance than the multitaper, which is consistent with the distribution of each estimator ($\chi^2_2$ vs.\ $\chi^2_{2K}$).}
    \label{fig:KeplerQ0}
\end{figure*}

The top panel of Figure \ref{fig:KeplerQ0} shows the multitaper power spectrum estimate $\hat{S}_{xx}(f)$ based on the $K = 6$ tapers shown in Figure \ref{fig:dpss}. For all time series shown in this paper, we subtract the sample mean $\bar{x} = (1/N) \sum_{t=0}^{N-1} x_n$---which is a zero-frequency term that can cause bias---before estimating the power spectrum. The light blue shading in Figure \ref{fig:KeplerQ0} shows the 95\% confidence interval, the calculation of which we will discuss in \S \ref{subsec:jk}. The vertical bar of the blue cross indicates the expected value of the jackknife variance estimate (see \S \ref{subsec:jk}), while the horizontal cross bar indicates the bandwidth $\tilde{\varpi}$. Green dots identify signals that exceed a significance threshold of $1-1/N = 0.99939$ according to the F-test for harmonic components (\S \ref{subsec:Ftest}). {The middle panel shows the p-value from the F-test conducted at each frequency; green dots indicate p-values below $1/N = 0.00061$, for which we reject the null hypothesis that the dataset contains no oscillation component at frequency $f$.}

Figure \ref{fig:KeplerQ0} demonstrates the variance reduction provided by multitapering: the 95\% confidence interval on $\hat{S}_{xx}(f)$ does not come close to encompassing the variability in $\hat{S}^{p}(f)$. For a time series $x_n$ consisting only of white noise, $\hat{S}_{xx}(f)$ forms a $\chi^2_{2\alpha}$-distributed stochastic process with $\alpha = \sum d^2_k(f) \lessapprox K$, which has stronger correlations across frequency and thus fewer large excursions than does the $\chi^2_2$ distributed periodogram $\hat{S}^p(f)$. {When equal weighting at all frequencies is used, $\alpha = K$. The distribution of the power spectrum estimator is derived in the \citet{brillinger75} textbook.}




\section{Multitaper extensions: Missing data, confidence intervals, and the F-test}
\label{sec:extensions}

Here we show how to compute multitaper power spectrum estimates from spacecraft data and identify statistically significant signals. In \S \ref{subsec:gaps} we demonstrate tapering for time series with gaps using the \citet{chave2019multitaper} method. We compare the spectral windows produced by the Chave multitaper method and the generalized Lomb-Scargle periodogram in \S \ref{subsec:windows}. Finally, we demonstrate two methods for assessing the statistical significance of oscillatory signals: jackknife confidence intervals (\S \ref{subsec:jk}) and the F-test for harmonic components (\S \ref{subsec:Ftest}).

\subsection{Missing-data multitaper estimator}
\label{subsec:gaps}

Almost all time-series observers have to contend with missing data due to instrument downtime or poor environmental conditions. In space-based astronomy, pieces of dust may reflect sunlight into the telescope barrel, 
yielding sudden background flux increases that cannot always be modeled out of the light curves. Furthermore, {\it Kepler} and K2 observations were interrupted once per month during data downlink (see \citet{garcia11} for a description of all types of {\it Kepler} observing interruptions). The TESS reaction-wheel momentum dumps that occur every 2.5~days increase the pointing jitter to $\sim 1$' for a few minutes at a time, creating low-quality light curve segments that some observers clip out \citep[e.g.][]{plachy21}. 

\citet{chave2019multitaper} extended the notion of maximally bandlimited orthogonal sequences introduced by \citet{s78} to those sampled on a grid where there is missing data. Such sequences are defined only at timestamps for which data are available. As in \S \ref{sec:multitaper}, we set the bandwidth $\varpi$ for the desired spectral window, and let the sequences be defined on the length $N$ time grid $\{t_n\}_{n=0}^{N-1}$ with unit sampling ($\Delta t = 1$) except where there are gaps---for example, $N = 7$ and $\{t_n\} = \{0, 1, 2, 3, 7, 8, 9\}$. The missing-data Slepian sequences (MDSS) solve an eigenvalue problem similar to Equation \ref{eq:slepian}:
\begin{equation} \label{eq:mdslep} 
  \lambda_k \dpss{k}{n} =
  \sum_{m=0}^{N-1} \frac{\sin 2\pi \varpi (t_n - t_m) }{\pi (t_n - t_m)} \dpss{k}{m},
\end{equation}
except on the missing-data time grid $\{t_n\}_{n=0}^{N-1}$ instead of the consecutive integers $0,\ldots,N-1$. The MDSS form an orthonormal set on $\{t_n\}_{n=0}^{N-1}$ and can be used as tapers to produce a missing-data multitaper (MDMT) power spectrum estimate:
\begin{equation}
    \hat{S}_{xx}(f) = \frac{1}{K} \sum_{k=0}^{K-1} |\mathcal{F} \{ w^{(k)}_t x_n \}|^2,
    \label{eq:mdpowspec}
\end{equation}
where $\mathcal{F}$ denotes the nonuniform fast Fourier transform (NFFT).\footnote{See \citet{keiner2009using} and \citet{barnett19} for a description of the NFFT algorithm.} Although it is possible to use frequency-dependent weighting (Equations \ref{eq:Eigenspec} and \ref{eq:weights}) for missing-data multitaper, it often yields power spectrum estimates with too few degrees of freedom for the F-test (\S \ref{subsec:Ftest}), so we do not use it here. We note that \citet{springford20} presented a multitaper method for {\it Kepler} data that relies on interpolating the DPSS tapers. However, interpolated DPSS are in general not orthogonal on the missing-data time grid, so the independence of the $\hat{S}^{(k)}_{xx}$ is lost and jackknife confidence intervals cannot be computed.

\begin{figure*}
    \centering
    \includegraphics[width=0.7\textwidth]{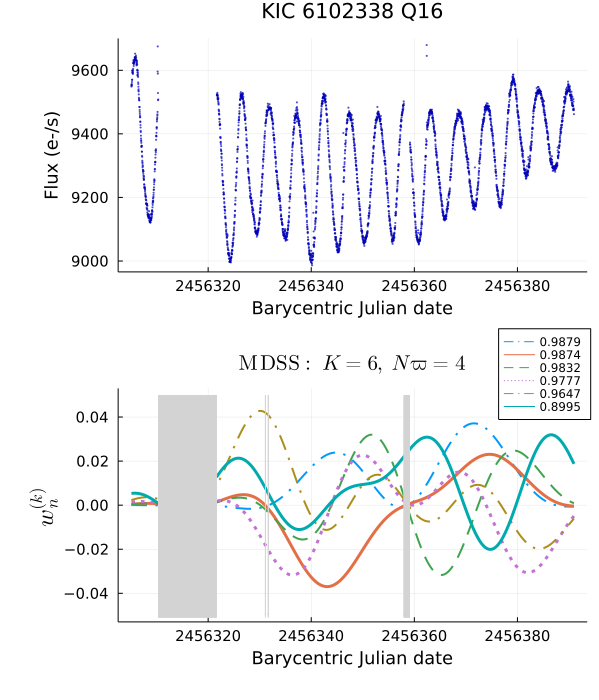}
    \caption{{Top:} PDCSAP fluxes from the Kepler Q16 observations of KIC~6102338. {Bottom:} Missing-data Slepian sequences $w^{(k)}_t$ for the 3534 observations plotted above, constructed with $N\varpi = 4$ and $K = 6$. The legend shows the spectral concentration of each $W^{(k)}(f)$. Gray shading highlights timestamps with missing data.}
    \label{fig:KeplerQ16flux_tapers}
\end{figure*}

Figure \ref{fig:KeplerQ16flux_tapers} shows the PDCSAP fluxes from the {\it Kepler} Q16 observations of KIC~6102338 (top) and a set of MDSS tapers $\dpss{k}{n}$ made for those observations (bottom), for which $N = 3534$, $N\varpi = 4$ ($\tilde{\varpi} = 0.05539$~day$^{-1}$), and $K = 6$. Gaps are highlighted by gray shading. Note that all $w_n^{(k)}$ approach zero at both edges of the large gap {near the beginning of the time series}: this behavior suppresses spectral leakage by removing the false high-frequency signal created by the sudden cutoffs in $x_n$ at the gap edges. (In fact, given the near-zero weighting for the earliest data, it would be reasonable to limit the analysis to BJD$> 2456322$.) Regardless of $N$ and $N\varpi$, {the two DPSS/MDSS tapers with the highest spectral concentrations}, $w_n^{(0)}$ and $w_n^{(1)}$, approach zero at the beginning and end of $x_n$ plus the edges of major gaps. We also see $w_n^{(0)}, w_n^{(1)}, w_n^{(2)}, w_n^{(3)}$, and $w_n^{(4)}$ approaching zero at the edges of the smaller gap that begins at BJD 2456357.96. 

The {top} panel of Figure \ref{fig:KeplerQ16spec} shows the missing-data multitaper power spectrum estimate $\hat{S}_{xx}(f)$ from the KIC~6102338 Q16 data and tapers shown in Figure \ref{fig:KeplerQ16flux_tapers}. Our choice of $N\varpi = 4$ for both Q0 and Q16 yields a narrower bandwidth $\varpi$ for Q16, which has more observations ($N = 3534$ for Q16 vs.\ $N = 1639$ for Q0). {The middle panel shows the p-value associated with the F statistic as a function of frequency (\S \ref{subsec:Ftest}).} As in Figure \ref{fig:KeplerQ0}, the green dots show oscillations significant at or above the $1-1/N$ level identified by the F-test. 
The {bottom panel} shows the Lomb-Scargle periodogram (gray) overplotted on the multitaper 95\% confidence interval (orange). Figure \ref{fig:KeplerQ16spec} highlights the variance suppression ability of the missing-data multitaper: the 95\% confidence interval does not bound the vertical excursions of the Lomb-Scargle periodogram.

\begin{figure*}
    \centering
    \includegraphics[width=0.7\textwidth]{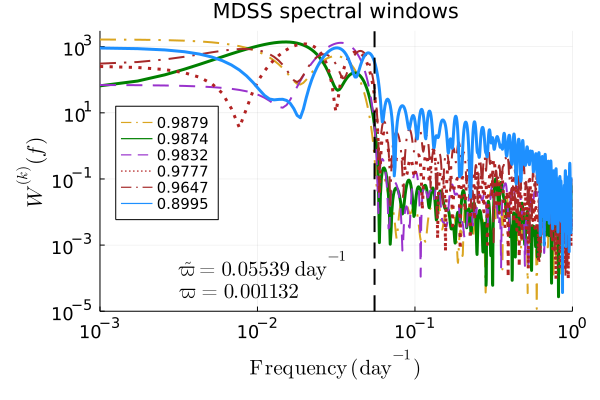}
    \caption{Spectral windows $W^{(k)}(f)$ from the gpss tapers applied to the {\it Kepler} Q16 observations of KIC~6102338 (Figure \ref{fig:KeplerQ16flux_tapers}). The legend indicates the spectral concentrations and the vertical dashed line marks the bandwidth.}
    \label{fig:KeplerQ16windows}
\end{figure*}

\begin{figure*}
    \centering
    \includegraphics[width=0.7\textwidth]{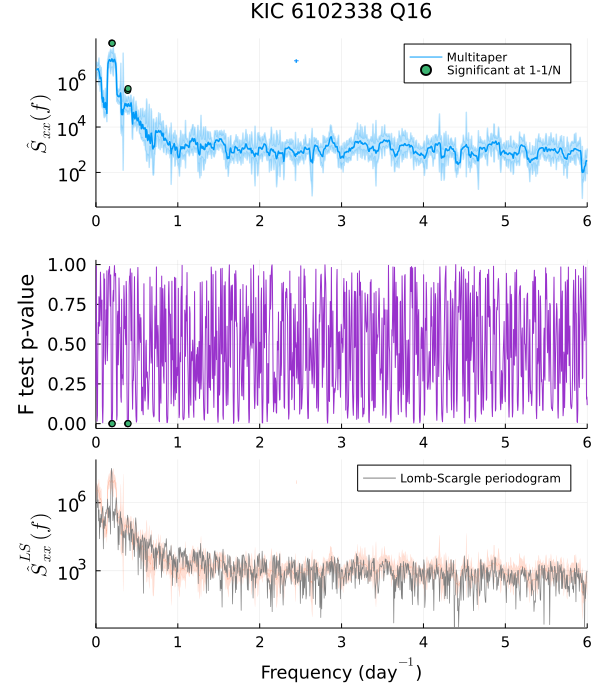}
    \caption{{Top:} The missing-data multitaper power spectrum estimate calculated using the tapers in Figure \ref{fig:KeplerQ16flux_tapers}. The shaded blue region shows the 95\% confidence interval, while the green dots show oscillatory signals that are significant at the 99.9\% level according to the F-test. The vertical crossbar has height $E \{ \widehat{\Var} \hat{S}_{xx}(f) \} \}$, while the horizontal crossbar has width $\tilde{\varpi} = 0.05539$~day$^{-1}$. {Middle:} The p-value given by the F-test for harmonic components, which is performed at each frequency (\S \ref{subsec:Ftest}). Green dots mark the frequencies for which $p < 1/N$. {Bottom:} The Lomb-Scargle periodogram (gray) overplotted on the multitaper 95\% confidence intervals (orange). The Lomb-Scargle periodogram has larger variance than the multitaper, which is consistent with the distribution of each estimator ($\chi^2_2$ vs.\ $\chi^2_{2K}$).}
    \label{fig:KeplerQ16spec}
\end{figure*}

The $k^{\rm th}$ MDSS spectral window $W^{(k)}(f)$ can be calculated using Equation \ref{eq:spectralwindow}, replacing the standard fast Fourier transform with the nonuniform fast Fourier transform \citep{barnett19, barnett20}. Figure \ref{fig:KeplerQ16windows} shows the positive half of each $W^{(k)}(f)$ calculated from the $w_n^{(k)}$ pictured in Figure \ref{fig:KeplerQ16flux_tapers} (as in the DPSS, the MDSS spectral windows are symmetric about zero). All $W^{(k)}(f)$ are orthonormal over the frequency range $-1/2 < f < 1/2$ and orthogonal on $-\varpi < f < \varpi$. Importantly, the spectral window of the missing-data multitaper estimator can be written analytically in terms of the individual $W^{(k)}(f)$'s. If one replaces the adaptive weights $d^2(f)$ (Equation \ref{eq:weights}) with $1/K$, one obtains the following expression for the missing-data spectral window:
\begin{equation}
W(f) = \frac{1}{K} \sum_{k=0}^{K-1} W^{(k)}(f).
\label{eq:mdmtwin}
\end{equation}

The MDSS spectral window delivers a lower dynamic range than the DPSS spectral window, as we can see by comparing the two versions of $W^{(0)}(f)$ in Figures \ref{fig:dpss} and \ref{fig:KeplerQ16spec}: the DPSS $W^{(0)}(f)$ drops by over 10 orders of magnitude over the frequency range $0 < f < \varpi$, while the MDSS $W^{(0)}(f)$ drops by less than five orders of magnitude over $0 < f < \varpi$. In general, missing-data tapers have lower spectral concentrations than classical tapers, leading to more spectral leakage. If the dataset is only missing a few scattered data points, as in the KIC~6102338 Q0 data, a bias-averse observer may wish to fill in the missing data and use the classical multitaper method instead of the \citet{chave2019multitaper} missing-data method. 
However, despite their different bias reduction capabilities, the missing-data and standard multitaper power spectrum estimators share the formal property that {all power spectrum peaks have the same rectangular shape as the spectral window}, which is in general not true for the Lomb-Scargle periodogram (see \citet[Appendix D]{scargle1982studies} and \S \ref{subsec:windows} on pseudowindowing).

\subsection{Spectral window comparison: Lomb-Scargle vs.\ missing-data multitaper}
\label{subsec:windows}

From Equation \ref{eq:expectedspectrumestimate} we know that the spectral window contributes bias by convolution with the true power spectrum, the characteristics of which require some special consideration when there are missing data. Figure 3 of \citet{scargle1982studies} shows dramatic comparisons between spectral windows from time series with unequal observing cadence vs.\ unit observing cadence. \citet{scargle1982studies} recommends the following heuristic to approximate the spectral window, or \emph{pseudowindow}, at each frequency: compute the Lomb-Scargle periodogram of a sinusoid with frequency $f$ (where $f$ is far from zero) on the set of observation times $\{t_n\}_{n = 0}^{N-1}$. Plotting this pseudowindow, which is convolved with the true process power spectrum in the neighborhood of frequency $f$, gives an idea of the spectral leakage.

\begin{figure*}
\centering
\includegraphics[width=0.8\textwidth]{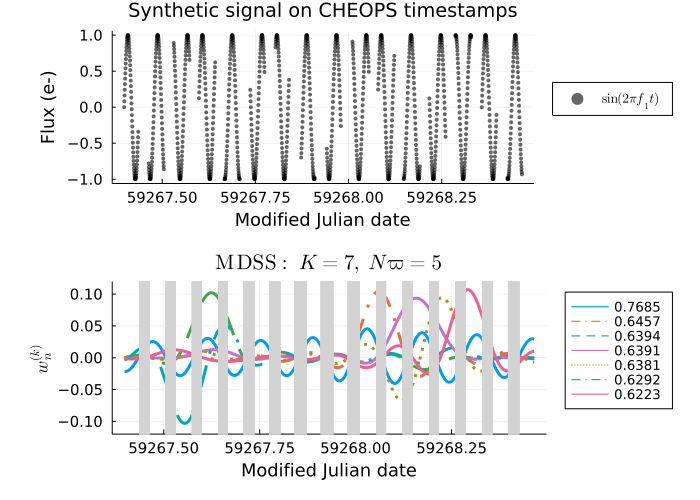} \\
\caption{{Top:} Synthetic signal $x_n = \sin (2 \pi f_1 t)$, where $f_1 = 25$~day$^{-1}$, placed on the CHEOPS 55 Cnc observation timestamps from MJD 59267.398-59268.466. {Bottom:} MDSS tapers with $N \varpi = 5$, $K = 7$, and $\tilde{\varpi} = 7.61$~day$^{-1}$ constructed for the CHEOPS observations. Gray shading highlights timestamps with missing data.}
\label{fig:CHEOPStapers}
\end{figure*}

As an example of pseudowindowing, we take time stamps from CHEOPS observations of 55~Cnc obtained on 22-23 February 2021 (MJD 59267.398-59268.466, PI: B.\ Demory). We construct two synthetic signals of the form $x_n = \sin (2 \pi f_0 t)$ with $f_0 = 25$~day$^{-1}$ and $f_0 = 80$~day$^{-1}$ on the grid of observation times. The top panel of Figure \ref{fig:CHEOPStapers} shows the synthetic signal with $f_0 = 25$~day$^{-1}$. From each signal, we compute a Lomb-Scargle periodogram and MDMT power spectrum estimate. The bottom panel shows the MDSS computed for $N\varpi = 5$, $K = 7$, $\tilde{\varpi} = 7.61$~day$^{-1}$, with the legend showing the spectral concentrations $\lambda^{(k)}$. We selected $N\varpi = 5$, rather than $N\varpi = 4$ as for the {\it Kepler} Q16 data, in order to widen the bandwidth and increase the spectral concentrations: with $N\varpi = 4$, the CHEOPS timestamps give $\lambda^{(0)} = 0.643$ for $W^{(0)}(f)$, compared with $\lambda^{(0)} = 0.769$ for $N\varpi = 5$. For a given $N\varpi$, time series with more gaps (e.g.\ CHEOPS) will have lower MDSS spectral concentrations than time series with fewer gaps (e.g.\ {\it Kepler} Q16).

\begin{figure*}
\centering
\begin{tabular}{c}
\includegraphics[width=0.95\textwidth]{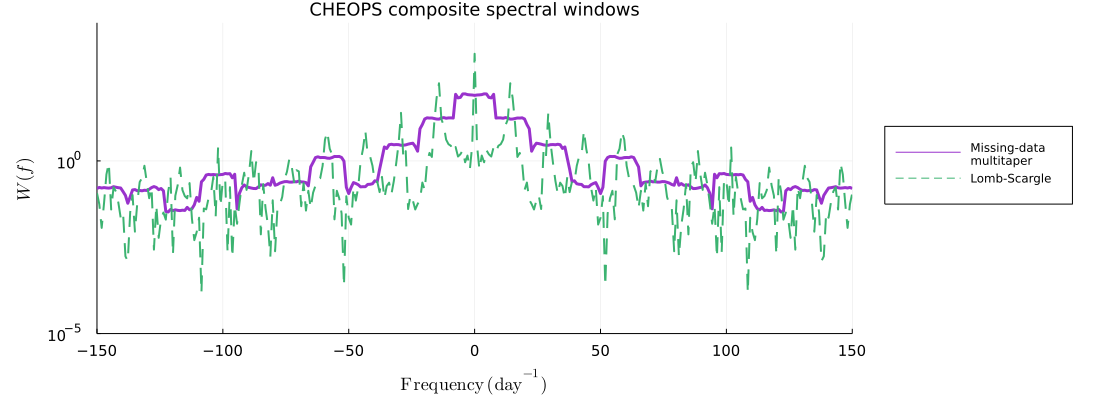} \\
\includegraphics[width=0.95\textwidth]{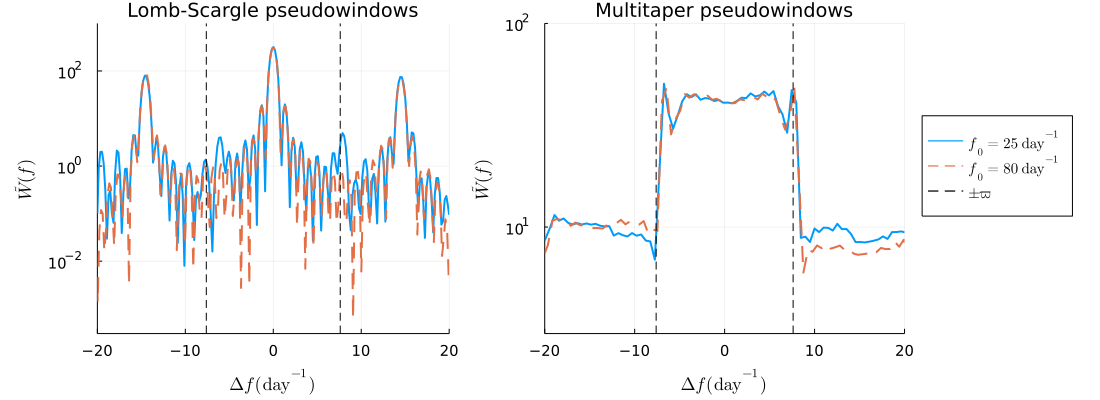}
\end{tabular}
\caption{{Top:} Multitaper spectral window $W(f)$ (purple) and Lomb-Scargle proxy spectral window $W^{LS}(f)$ (green dashed). {Bottom left:} Lomb-Scargle pseudowindows for $f_0 = 25$~day$^{-1}$ (blue) and $f_0 = 80$~day$^{-1}$ (orange dashed). {Bottom right:} Missing-data multitaper pseudowindows for $f_0 = 25$~day$^{-1}$ (blue) and $f_0 = 80$~day$^{-1}$ (orange dashed).}
\label{fig:pseudowindows}
\end{figure*}

The top panel of Figure \ref{fig:pseudowindows} compares the spectral window of the MDMT, $W(f)$ (purple), with a proxy for the zero-frequency spectral window of the Lomb-Scargle periodogram $W^{LS}(f)$ (green dashed). We calculate $W^{LS}(f)$ using the non-uniform fast Fourier transform:
\begin{equation}
    W^{LS}(f) = \left| \mathcal{F} \left \{ \frac{\operatorname{I}_N(t)}{\sqrt{N}} \right \} \right|^2,
    \label{eq:lsspecwin}
\end{equation}
where the indicator function is $\operatorname{I}_N(t) = 1$ for $t = t_0, ..., t_{N-1}$ and zero otherwise \citep{vanderplas18}. Our definition of $W^{LS}(f)$ shows the shape of the zero-frequency Lomb-Scargle spectral window but uses the power spectral density normalization of the multitaper. One immediately obvious feature of $W(f)$ from the missing-data multitapers is the sidelobe suppression: while leakage in the Lomb-Scargle spectral window manifests as a sawtooth pattern in which the highest sidelobes at $f = \pm 15$~day$^{-1}$ have 14\% of the amplitude of the main lobe (the central peak of the spectral window), the MDMT lacks the $\pm 15, 30, 45$~day$^{-1}$ sidelobes. The highest secondary peak in the MDMT spectral window at $f = \pm 57$~day$^{-1}$ has only 1.4\% the amplitude of the main lobe, greatly decreasing the possibility of misinterpreting an observing cadence artifact as a true signal \citep[e.g.][Ch.\ 3]{koopmans95}.

The bottom two panels of Figure \ref{fig:pseudowindows} show the pseudowindows centered at frequencies $f_0 =  25$~day$^{-1}$ (blue solid line) and $f_0 = 80$~day$^{-1}$ (orange dashed line) in the Lomb-Scargle periodogram (left) and MDMT (right). In both plots, the x-axis variable is $\Delta f = f - f_0$ and the pseudowindows are zoomed in to $|\Delta f| \leq 20$~day$^{-1}$ in order to focus on the main lobe and the first pair of Lomb-Scargle sidelobes. While the two Lomb-Scargle pseudowindows have similar shapes to $W^{LS}(f)$ and to each other, they are not identical: at $\Delta f = 8$~day$^{-1}$ they differ by more than a factor of three, while at $\Delta f = -20$~day$^{-1}$ they differ by an order of magnitude. In contrast, the MDMT behaves almost as well in reality as it does in theory: the most significant difference between the $f_0 = (25, 80)$~day$^{-1}$ pseudowindows is 20\%. The CHEOPS time series shows the value of reshaping the spectral window with the MDSS, even when the gaps are long and numerous.

\subsection{\label{sec:falsepositives} Spurious peaks} 

In this section we briefly summarize a theoretical result concerning significant peaks in nonparametric power spectrum estimators \citep{th14}. Restricting ourselves to the unit sampling ($\Delta t = 1$) case for the moment, we know from \S \ref{subsec:classicalmpse} that periodograms have a $\chi^2_2$ or exponential distribution and that multitaper power spectra have a $\chi^2_{2K}$ distribution (when frequency-dependent weighting is used, the distribution is $\chi^2_{\alpha}$ where $\alpha = 2\sum_{k=1}^K d^2_k(f)$ at frequency $f$). One can then assign significance to large peaks in a power spectrum estimate (after one has subtracted the baseline broadband spectrum, or \emph{prewhitened}) using the appropriate $\chi^2$ quantile. The contribution in \cite{th14} was to show that the tendency of the power spectrum estimate to spuriously upcross a high threshold, $z$, depends on the second derivative of the serial correlation (dependency across frequency) of the estimator. These upcrossings occur at a rate of
\begin{equation}
r = \sqrt{\frac{\pi z}{3}} \, e^{-z}
\label{eq:per_upcrossing}
\end{equation}
for the periodogram and 
\begin{equation}
r_{MT} = \frac{\sqrt{z}}{\pi} \cdot \frac{1}{\alpha} \cdot \frac{\alpha^{\alpha}}{\Gamma(\alpha)} \, z^{\alpha - 1} e^{-\alpha z}
\label{eq:mt_upcrossing}
\end{equation}
for the multitaper estimator (upcrossing rates are expressed per normalized frequency unit, or Rayleigh resolution).

Equations \ref{eq:per_upcrossing} and \ref{eq:mt_upcrossing} confirm that multitaper produces fewer spurious upcrossings per significance threshold by a factor of almost $2N\varpi$. This statistical fact is further exemplified in Thomson and Haley (2014) with numerical experiments on synthetic white-noise time series
and proton density solar-wind data from the Advanced Composition Explorer spacecraft (their Table 2). \citet{th14}, Figure 2a shows that when the time-bandwidth product $N\varpi$ is chosen as 5 (and, in fact this holds for all $N\varpi > 1$), the multitaper power spectrum gives approximately five times fewer spurious peaks than the periodogram (with white noise, all peaks are spurious.) As an aside, we might just find significant telescope time savings when faced with a, say, fivefold decrease in spurious oscillation detections as a consequence of switching from periodograms to the multitaper method!

\subsection{Calculating confidence intervals}
\label{subsec:jk}

With multitaper, we can go further than just estimating a power spectrum---we can also compute confidence intervals. In particular, we are interested in a confidence interval for the logarithm of the power spectrum. Since the $\log \hat{S}^{(k)}_{xx}(f)$ are approximately independent, they can be used to estimate variance. The \emph{delete one jackknife} estimator of \citet{efronstein81, t90B, thomson91, T94jk} is less biased than the sample variance of the individual eigenspectra \citep{cressie81} and is closer to the true value $\Var \log \hat{S}_{xx}(f) \}$ even for small $K$.





To jackknife, one computes $K$ delete-one estimators of the log-eigenspectrum $\log \hat{S}_{xx}(f)$. The logarithmic transformation ensures that the set of jackknife pseudovalues, which we will define shortly, is approximately normally distributed at each frequency \citep[\S 5.3]{Miller74}. For the simple average multitaper estimator, which we use for time series with missing data (Equation \ref{eq:mdpowspec}), the delete-one log-eigenspectra are
\begin{equation}
\log \hat{S}^{\setminus m}_{xx}(f) = \log \left[ \frac{1}{K-1} \sum_{\substack{k = 0 \\ k \neq m}}^{K-1} \hat{S}_{xx}^{(k)}(f) \right]
\label{eq:simplejackknife}
\end{equation}
The jackknifed variance estimate is based on the sample variance of quantities termed \emph{pseudovalues}, and not the delete-one log-eigenspectra themselves. The pseudovalues are defined as 
\begin{equation}\label{eq:pseudo}
p_m(f) = K\log \hat{S}_{xx}(f) - (K-1) \log \hat{S}^{\setminus m}_{xx}(f)
\end{equation}
and can be treated as if they come from a normally distributed sample \citep{cressie81}, which means their variance is Student $t-$distributed with $K-1$ degrees of freedom. Computing the variance of the $p_m(f)$'s gives the following log power spectrum variance estimate:
\begin{equation}\label{eq:jkvar}
\widehat{\Var} \log \hat{S}_{xx}(f) \} = \frac{1}{K(K-1)} \sum_{m=0}^{K-1} \left[ p_m(f)- \frac{1}{K} \sum_{k=0}^{K-1} p_m(f) \right]^2.
\end{equation}
We have constructed the $100(1-\alpha)\%$ confidence intervals according to 
\begin{eqnarray} \label{eq:JKCI}
& \hat{S}_{xx}(f) \: \exp \left \{\mathfrak{t}_{K-1}(\alpha/2) \sqrt{\widehat{\Var} \log \hat{S}_{xx}(f) \}} \right \} < S(f) \nonumber \\ & < \hat{S}_{xx}(f) \: \exp \left \{\mathfrak{t}_{K-1}(1-\alpha/2) \sqrt{\widehat{\Var} \log \hat{S}_{xx}(f) \}} \right\}
\end{eqnarray}
as in \citet[Eq. (2.49)]{TC91}, where $\mathfrak{t}_{K-1}$ is the Student-$t$ distribution with $K-1$ degrees of freedom and $\alpha$ is the significance level. 

Figures \ref{fig:KeplerQ0} and \ref{fig:KeplerQ16spec} both use light blue shading to show jackknife 95\% confidence intervals on the multitaper power spectrum estimators. The confidence intervals provide guidance for interpreting the results of the F-test for harmonic components (\S \ref{subsec:Ftest}): a power spectrum peak may be quite high, but if the confidence interval in and around that peak is wide, the F-test p-value might not be statistically significant. The combination of confidence intervals and the F-test allows the observer to take into account the quality of information at a particular frequency before deciding if an oscillation at that frequency is significant. Confidence intervals are also important when the science goal requires not only identifying oscillations, but precisely estimating their power (see example in \S \ref{sec:kepler_diffrot}).

\subsection{F-test for harmonic components}
\label{subsec:Ftest}

Adjacent to the topic of power spectral estimation, but separate from it, 
is harmonic analysis, at which the multitaper technique also excels. {Harmonic analysis, which is the detection of oscillations, is the typical astronomical application of  periodograms. For example, radial velocity exoplanet hunters will begin often begin an analysis with a generalized Lomb-Scargle periodogram, then fit a time-domain model (for example, a Keplerian for a planet or a Gaussian process for rotation) using the period associated with the highest periodogram peak to initialize the fit. The F-test is the harmonic analysis tool associated with the multitaper technique.} 
We refer the reader interested in technical details to the original source paper \citep{thomson1982spectrum} and give a basic presentation here, with the relevant extensions for time series with missing data. {For an accessible textbook overview of the F-test, see \citet[Chapter 5]{mudelsee14}.}

With the Lomb-Scargle periodogram, the usual procedure is to assume the noise is white and use either bootstrapping \citep{cumming04} or extreme-value theory \citep{baluev08} to calculate false alarm probabilities (FAPs) that are constant as a function of frequency. However, both turbulence and signal persistence in a bolometer produce red noise, such that the continuum of $\hat{S}_{xx}(f)$ underlying any oscillations has more power at low frequencies than high frequencies. Although \cite{delisle2020} provide a generalization of the Lomb-Scargle periodogram for noise with arbitrary covariance structure, their method requires the user to estimate a correlation time in order to approximate the covariance matrix. Given that the true correlation time is always unknown, \citet{delisle2020} recommend (1) examining multiple periodograms computed using different correlation times and (2) using a likelihood maximization to calculate a more realistic covariance matrix. It would be useful to be able to judge the significance of a power spectrum peak without needing to make any assumptions about the noise.
{In addition, a FAP estimate that is robust against large dynamic ranges in $\hat{S}_{xx}(f)$ would reduce the need to iteratively subtract sinusoids or Keplerian models from the time series and examine periodograms of the residuals \citep[e.g.][]{reinhold13, bourrier18}, a practice that can be error-prone given uncertainties on the correct sinusoid frequencies. }

\citet{thomson1982spectrum} gives an effective test for the presence of oscillatory components (i.e.\ purely sinusoidal signals) in a multitaper power spectrum. Thomson's F-test generalizes easily to the missing-data case \citep{chave2019multitaper}. The test statistic is 
\begin{equation}\label{eq:Fstat}
F(f) = \frac{(K-1) \; |\hat{\mu}(f)|^2 \; \sum_{k=0}^{K-1} [\tilde{w}^{(k)}(0)]^2}{\sum_{k = 0}^{K-1} \left| \tilde{x}^{(k)}(f) - \hat{\mu}(f) \: \tilde{w}^{(k)}(0) \right|^2},
\end{equation}
where $\tilde{x}^{(k)}(f)$, the eigencoefficients, are given by the complex-valued Fourier transform of the tapered time series
\begin{align}
    \tilde{x}^{(k)}(f) = \sum_{t=0}^{N-1} w^{(k)}_t x_n  e^{-2 \pi i f t} \; \; &(\operatorname{unit} \; \operatorname{sampling}), \label{eq:uniteigencoef}\\
    \tilde{x}^{(k)}(f) = \mathcal{F} \{ w^{(k)}_t x_n \} \; \; &(\operatorname{missing} \; \operatorname{data}) \label{eq:eigencoef}
\end{align}
and $\tilde{w}^{k}(0)$ is the Fourier transform of taper $k$, evaluated at zero frequency:
\begin{align}
    \tilde{w}^{(k)}(0) = \sum_{t=0}^{N-1} w^{(k)}_t\; \; &(\operatorname{unit} \; \operatorname{sampling}), \label{eq:unittapertrans}\\
    \tilde{w}^{(k)}(0) = \mathcal{F} \{ w^{(k)}_t \} \; \; &(\operatorname{missing} \; \operatorname{data}).
    \label{eq:tapertrans}
\end{align}
The estimated complex oscillation component, before squaring, is
\begin{equation}\label{eq:muhat}
\hat{\mu}(f) = \frac{\sum_{k=0}^{K-1} \tilde{w}^{(k)}(f) \; \tilde{x}^{(k)}(f)}{\sum_{k=0}^{K-1} \left[ \tilde{w}^{(k)}(f) \right]^2}.
\end{equation}
The test statistic (Equation \ref{eq:Fstat}) has an $F-$distribution where the numerator and denominator have $2$ and $2K-2$ degrees of freedom, respectively. At frequencies where the F-test p-value is small, there is support for the hypothesis that the time series records an oscillation.\footnote{Formally, the hypothesis being tested at each frequency gridpoint $f_j$ is that if we use the \citet{munk64} representation for the deterministic, first-moment part of the power spectrum, $E \{ dX(f) \} = \sum_j C_j \delta (f - f_j) df\}$ (where $dX$ is defined by the Cram\'{e}r spectral representation, $x_n = \int_{-1/2}^{1/2} \exp \left[ -2 \pi i f t\right] dX (f)$, and $C_j$ is the complex Fourier coefficient at frequency gridpoint $j$), the value of $C_j$ is nonzero \citep{thomson90}. The power spectrum is also assumed to have a non-deterministic noise component given by the second moment of $dX$. Since $S(f) df = E \{ |dX(f)|^2 \}$, each nonzero $C_j$ corresponds to a delta function in the true power spectrum, which manifests as a narrow, rectangular peak in the multitaper power spectrum estimate due to Equation \ref{eq:expectedspectrumestimate}. Physically, delta functions in the power spectrum correspond to sinusoids.} We choose a stringent F-test p-value cutoff of $1/N$ \citep{thomson90}, for which the expected number of random sampling-based fluctuations that reach our detection threshold is one per power spectrum estimate. The test statistic is to be interpreted as the estimated power at frequency $f$ compared with the power one obtains when a sinusoid with frequency $f$ is removed from the time series. Some detected oscillatory components may not coincide with large peaks in the power spectrum, especially when signal to noise ratios are low.

While power spectrum peaks spaced closer than twice the bandwidth ($\Delta f < 2N\varpi$) may be difficult to distinguish if classical multitaper and MDMT estimators are the only tools available, careful inspection of the F-test statistic reveals that its fundamental frequency unit is the Rayleigh resolution (Equation \ref{eq:Rayleigh}). Thus the F-test can distinguish signals with small enough frequency separation that {their rectangular power spectrum peaks overlap} in a plot of $\hat{S}_{xx}(f)$. It can also pick out long-period signals with $f \leq N \varpi$ that are not visually distinct from zero frequency as long as those signals have $f > 2 \mathcal{R}$, {meaning they are formally resolvable according to the Rayleigh criterion. The multitaper--F-test combination therefore has the same frequency resolution as Bayesian, least-squares, and generalized Lomb-Scargle periodograms, and is suitable for detecting multiple periodicities \citep[][see \S \ref{sec:scallop}, in which we identify 10 distinct oscillation modes in a ``scallop shell'' light curve]{vautard92, robertson98, zhou07}.} 
{Furthermore, the F-test is valid no matter the type of noise in the dataset, as demonstrated analytically by \citet{thomson1982spectrum, thomson90} and computationally for synthetic and observed seismic datasets by \citet{park87}, \citet{lees95}, and \citet{prieto07}, among others.} For accurate estimation of oscillation frequencies, oversampling of the frequency grid (or zero padding in the time domain) is recommended. Note, however, that oversampling does not add information to the power spectrum estimate; it merely interpolates the estimate.

Figures \ref{fig:KeplerQ0} and \ref{fig:KeplerQ16spec} both denote significant oscillations identified by the F-test with green dots. In Sections \ref{sec:kepler_diffrot} and \ref{sec:scallop}, we will follow the same convention and discuss the information provided by the F-test about each example dataset.

\section{KIC 6102338: Differential rotation, harmonics, and possible antisolar shear}
\label{sec:kepler_diffrot}

Our first missing-data multitaper demonstration uses the {\it Kepler} Q3 observations of KIC~6102338, an active M dwarf at a distance of 307~pc \citep{gaia3}. According to the study of stellar surface shear by \citet{reinhold13}, KIC 6102338 exhibits differential rotation with {primary period} $P_0 = 5.27$~days and {secondary period} $P_1 = 4.07$~days. {While angular rotation speed varies continuously as a function of latitude, it is common for two dominant periods to emerge in the photometry, as spots tend to favor certain latitudes \citep[e.g.][]{frasca11, frohlich12}.} \citet{reinhold13} searched for periodicities using iterative prewhitening: from a generalized Lomb-Scargle periodogram of each PDC-MAP light curve,\footnote{The PDC-MAP light curves are constructed so as to remove instrument systematics but preserve stellar variability; see \citet{stumpe12, smith12} for details.} they identified the period of maximum power $P_0$, subtracted a sinusoid with period $P_0$ from the light curve, and then computed a generalized Lomb-Scargle periodogram of the residuals. The procedure was repeated four times, and then $P_0$ plus the four periods of maximum power from each residual periodogram were used as initial guesses for the periods in a truncated Fourier series fit to the light curve:
\begin{equation}
    F = \sum_{i = 0}^4 A_i \sin \left( \frac{2 \pi t}{P_i} + \phi_i \right),
    \label{eq:reinholdmodel}
\end{equation}
where $F$ is the flux and the $P_i$ were allowed to vary. A detection of differential rotation was recorded when any of $P_1 \ldots P_4$ fell within 30\% of $P_0$. Here we will recover the same results as \citet{reinhold13}, but without iterative sinusoid subtraction. We will also use the {power at the} rotation harmonics to measure the direction of the stellar surface shear.

\subsection{Data and tapers}
\label{subsec:keplerdata}

\begin{figure*}
    \centering
    \includegraphics[width=0.7\textwidth]{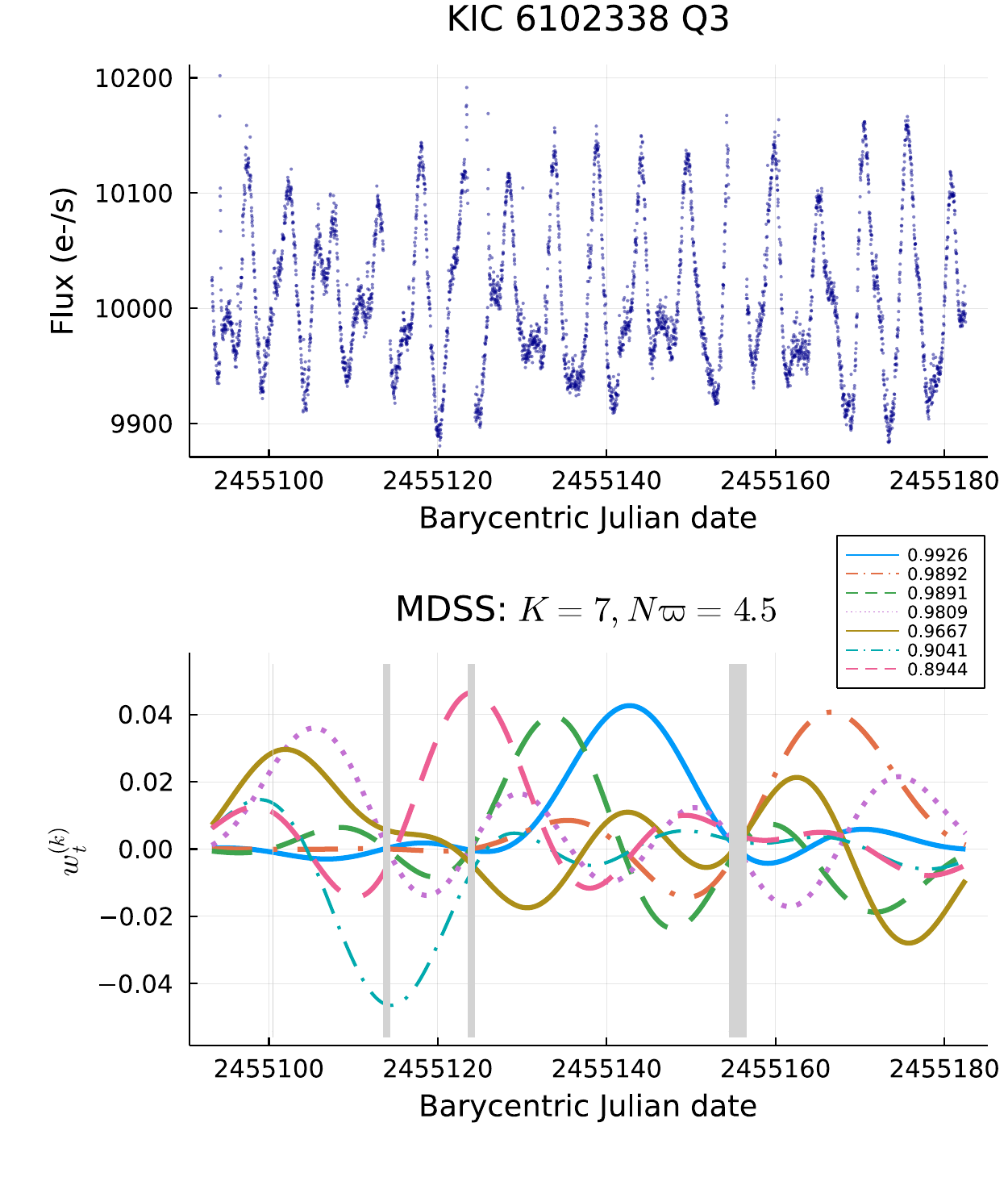}
    \caption{ {Top:} $x_n$, the {\it Kepler} Q3 PDC-MAP light curve of KIC 6102338 after removal of $3 \sigma$ outliers. {Bottom:} Missing-data Slepian sequences $\dpss{k}{t}$ applied to $x_n$. Gray shading highlights missing data and spectral concentrations $\lambda^{(k)}$ are given in the legend. For this set of tapers, $N\varpi = 4.5$ and $\tilde{\varpi} = 0.0533$~day$^{-1}$.}
    \label{fig:Kepler_tapers}
\end{figure*}

\citet{reinhold13} selected the {\it Kepler} Q3 data for their study because Q3 showed fewer instrument artifacts than Q0--Q2. {To perform a controlled comparison between multitaper and their iterative prewhitening procedure, we also base our analysis} on the KIC 6102338 Q3 long-cadence PDC-MAP light curve. We remove $>3\sigma$ outliers, which may be contaminated by stellar flares. The top panel of Figure \ref{fig:Kepler_tapers} shows the resulting light curve $x_n$ ($N = 4131$). The bottom panel of Figure \ref{fig:Kepler_tapers} shows the MDSS tapers with $N\varpi = 4.5$ and $K = 7$. We chose a wider bandwidth than we used for Q16 ($N\varpi = 4$, $K = 6$; \S \ref{subsec:gaps}) in order to improve variance suppression: with $N\varpi = 4.5$, we can safely include $w_n^{(6)}$, which has $\lambda^{(6)} = 0.894$---similar to $\lambda^{(5)}$ for Q16 (Figure \ref{fig:KeplerQ16flux_tapers}). By decreasing the variance of our power spectrum estimate, we reduce the uncertainty on the shape of each power spectrum peak, making it more likely that true oscillatory signals reach high statistical significance in the F-test. High dynamic range in $W(f)$ allows us to simultaneously detect the strong signals at the fundamental rotation frequencies and their weaker harmonics.


\subsection{Power spectrum estimate}
\label{subsec:keplerpowspec}

\begin{figure*}
    \centering
    \includegraphics[width=0.7\textwidth]{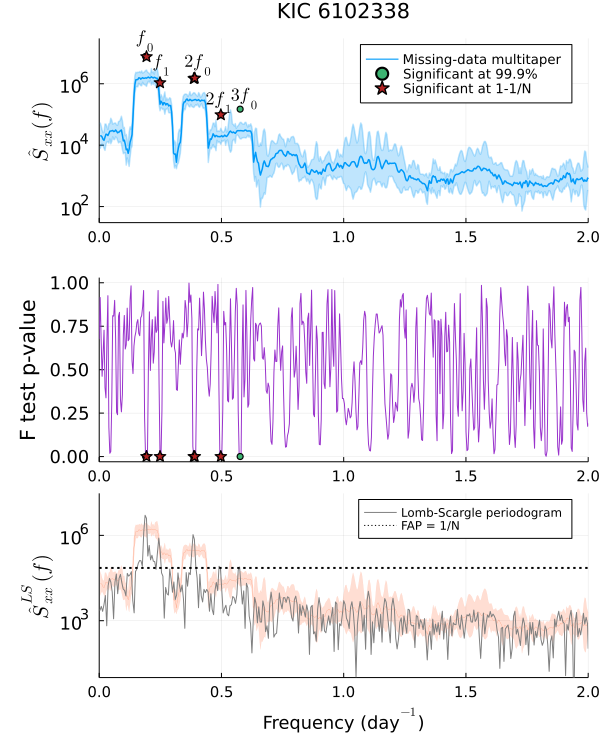}
    \caption{Power spectrum estimates from the time series $x_n$ plotted in Figure \ref{fig:Kepler_tapers}. {Top}: The blue line shows the missing-data multitaper with $w^{(k)}_t$ from Figure \ref{fig:Kepler_tapers}, with shaded 95\% confidence intervals. Red stars show signals with significance at least $1-1/N$ according to the F-test. The green circle shows an additional signal with $99.9$\% significance. Annotations give the frequencies of all significant oscillations. {Middle: The p-value associated with the F statistic at each frequency. The red stars and green circle have the same meaning as above. Bottom: The Lomb-Scargle periodogram plotted on top of the multitaper confidence intervals, with the $1/N$ false alarm level indicated by the dotted line.}}
    \label{fig:KeplerQ3powspec}
\end{figure*}

Figure \ref{fig:KeplerQ3powspec} shows the power spectrum estimates from the KIC 6102338 {\it Kepler} Q3 data. In the top panel, the blue line shows the MDMT with the 95\% confidence interval shaded in light blue. The red stars indicate oscillatory signals at the $1-1/N$ significance level, according to the F-test. The frequency of the strongest signal, $f_0 = 0.193$~day$^{-1}$, matches the primary rotation signal identified by \citet{boisse2012, reinhold13} to within $0.3 \mathcal{R}$, which is well below the fundamental resolution unit. We also detect an oscillation at $f_1 = 0.249$~day$^{-1}$, which is again within $0.3 \mathcal{R}$ of the secondary rotation signal reported by Reinhold et al. {The multitaper--F-test combination reproduces their differential rotation detection, but does not require any iterative optimization.}

The detections at $f_0$ and $f_1$ highlight the fact that the F-test can distinguish signals that are not separated by a full bandwidth: {visually, the two rectangluar power spectrum peaks at $f_0$ and $f_1$ overlap because they trace two oscillations that are separated by $\Delta f < 2 N \varpi$ (in this case, $2\tilde{\varpi} = 0.107$~day$^{-1}$, but $f_1 - f_0 = 0.0553$~day$^{-1}$). However, both corners of the $f_0$ peak and the right-hand corner of the $f_1$ peak are clearly visible. When examining multitaper power spectrum estimates, it's helpful to remember that power spectrum peaks are approximately rectangular; sharp corners usually belong to oscillatory signals.}

Figure \ref{fig:KeplerQ3powspec} shows more than just the two fundamental rotation frequencies. We also find significant oscillations at $2 f_0$ and $2 f_1$, the first harmonics of each rotation frequency. Furthermore, if we lower our statistical significance threshold from $1-1/N$ to 0.999, we pick up an oscillation at $3 f_0$, the second harmonic of the primary rotation signal (green dot). While detections of differential rotation without harmonics can be used to measure the {\it magnitude} of stellar surface shear, the harmonics provide an additional piece of information: the {\it direction} of the shear.

Measuring the power ratio of each first harmonic compared with its fundamental frequency,
\begin{equation}
    \mathcal{P}(f) = \frac{\hat{S}_{xx}(2 f)}{\hat{S}_{xx}(f)},
    \label{eq:powerratio}
\end{equation}
we find that $\mathcal{P}(f_0) = 0.194$, while $\mathcal{P}(f_1) = 0.0904$. Following \citet{reinhold15}, we assume that spot groups near the stellar equator produce stronger rotation harmonics than spot groups far from the equator. Our results suggest that $f_0$ is associated with regions nearer the equator than $f_1$, which means KIC 6102338 might exhibit antisolar differential rotation. Indeed, the possibility that $\mathcal{P}(f_1) > \mathcal{P}(f_0)$ does not fall within the 95\% confidence intervals on $\hat{S}_{xx}(f)$. However, the straightforward relationship between $\mathcal{P}(f)$ and spot latitude breaks down when there are large numbers of spot groups. Since the star's rapid rotation suggests that it is young \citep{barnes10} and therefore active \citep{mamajek08}, the spot patterns may be complex. Further information about the stellar surface (i.e.\ from Doppler imaging) would be useful for verifying the direction of the shear.

{Our analysis of the KIC 6102338 {\it Kepler} Q3 data demonstrates the value of accuracy and precision in power spectrum estimation: we cannot measure a differential rotation direction using Equation \ref{eq:powerratio} if (a) the confidence intervals are wide or (b) the bias is large. With its bias bounded by Equation \ref{eq:mtbias}, the MDMT can be constructed to meet a particular accuracy requirement in $\hat{S}_{xx}(f)$; the bias in $\hat{S}_{xx}(f)$ in Figure \ref{fig:KeplerQ3powspec} is 443~(e-/s)$^2$, which is $< 1\%$ of the power in our weakest signal under analysis (harmonic at $2 f_1$). The MDMT should be the preferred Fourier analysis tool for all science goals that require accurate, precise power estimates in addition to period searches.}

\section{EPIC 203354381: A complex rotator with a ``scallop-shell'' light curve}
\label{sec:scallop}

From K2 observations of the Upper Scorpius star-forming region, \citet{stauffer17} identified a new type of stellar variability that produces phased light curves with ``scallop shell'' shapes. Their hypothesis was that the complex light-curve morphologies result from orbiting material with the same period as the star's rotation (usually $< 1$~day), possibly due to ``slingshotting'' of gas in the stellar corona. \citet{stauffer18} followed up with the discovery of more scallop-shell stars in the Upper Sco, $\rho$~Oph, and Taurus star-forming regions. However, both papers confined their light-curve investigations to the time domain.

Here we use multitaper to analyze K2 Campaign 2 observations of EPIC~203354381, an Upper Sco scallop-shell star identified by \citet{stauffer18}. In \S \ref{subsec:epicdata} we discuss the time series and taper computation. In \S \ref{subsec:epicpower} we identify oscillation modes using the multitaper--F-text combination and discuss their possible physical origins.

\subsection{Data and tapers}
\label{subsec:epicdata}

\begin{figure*}
    \centering
    \includegraphics[width=0.8\textwidth]{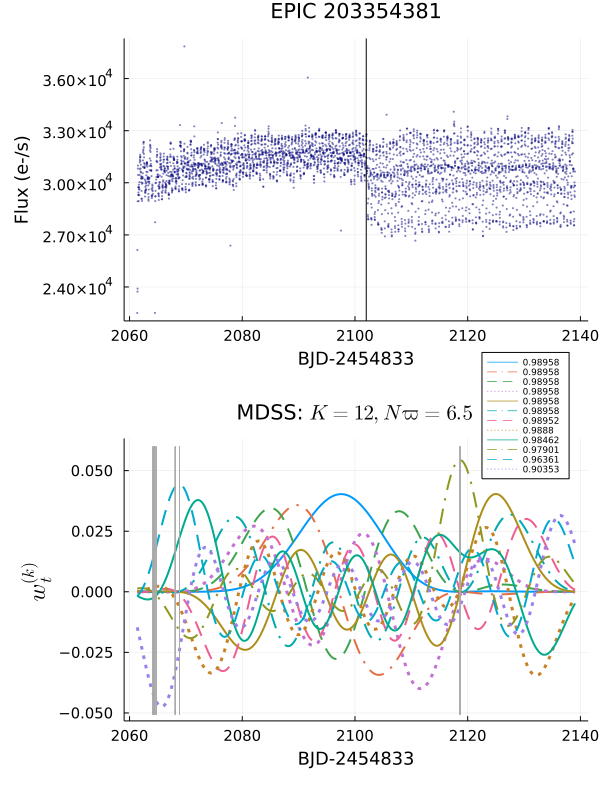}
    \caption{{Top:} Detrended K2 observations of scallop-shell star EPIC 203354381 with $>3\sigma$ outliers clipped. {Bottom:} Missing-data multitapers for $N \varpi = 6.5$, $K = 11$. Spectral concentrations are shown in the legend and gaps are highlighted in gray.}
    \label{fig:scallop_data}
\end{figure*}

{EPIC~203354381 was observed by the {\it Kepler} spacecraft in its K2 incarnation from 23 August 2014 to 10 November 2014. For our analysis, we use the detrended light curve produced by the EVEREST pipeline \citep{luger16} with $>3\sigma$ outliers clipped. We drop the first 50 observations, which record a stellar flare, from our analysis. The top panel of Figure \ref{fig:scallop_data} shows the light curve after outlier and flare rejection. At time $\operatorname{BJD} = 2456935$, marked with a vertical black line, the variability amplitude and light curve morphology suddenly change. We will present a multitaper power spectrum estimate from the entire dataset pictured in Figure \ref{fig:scallop_data}, plus separate power spectrum estimates from the early data (before $\operatorname{BJD} = 2456935$) and the late data (after $\operatorname{BJD} = 2456935$).

For the full dataset, which has $N = 3737$, we selected $N\varpi = 6.5$ and $K = 12$. Unlike the KIC 6102339 rotation signals, the EPIC 203354381 oscillatory modes are well separated in frequency, so we can increase the time-bandwidth product from $N\varpi = 4$, which we used in \S \ref{sec:multitaper} and \ref{sec:kepler_diffrot}, and still be able to distinguish the different modes in a plot of the power spectrum estimate. Recall, however, that the resolution of the F-test does not change with bandwidth. The bottom panel of Figure \ref{fig:scallop_data} shows the MDSS for $N\varpi = 6.5$ and $K = 11$, with the time series gaps highlighted in gray. Note the tendency of low-order MDSS (small $k$) to approach zero at the gap edges, a behavior which suppresses spurious high-frequency power produced by sudden observation stoppages (\S \ref{subsec:gaps}).}

\begin{figure*}
    \centering
    \includegraphics[width=0.7\textwidth]{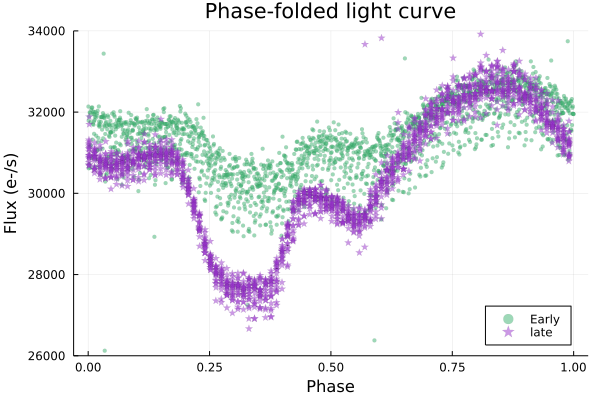}
    \caption{Phase-folded versions of the early (green circles) and late (purple stars) parts of the EPIC~203354381 light curve shown in the top panel of Figure \ref{fig:scallop_data}. The early and late parts of the dataset are before and after $\operatorname{BJD} = 2456935$, respectively.}
    \label{fig:scallop_folded}
\end{figure*}

{When we cut the time series in two using the vertical line in Figure \ref{fig:scallop_data} and analyze the early data ($N = 1944$) and late data ($N = 1793$) separately, we revert to $N\varpi = 4$, $K = 6$. The abrupt change in variability amplitude is highlighted in Figure \ref{fig:scallop_folded}, which shows phase-folded versions of the early and late data subsets using the rotation period inferred from the multitaper power spectrum estimates (see \S \ref{subsec:epicpower} below). The phased light curve shapes are similar, but the ``sharpness'' of the late-data light curve suggests that more high-frequency modes are present in the late data than the early data. We will see this prediction borne out in the next section.
}

\subsection{Power spectrum estimates}
\label{subsec:epicpower}

\begin{figure*}
    \centering
    \includegraphics[width=0.7\textwidth]{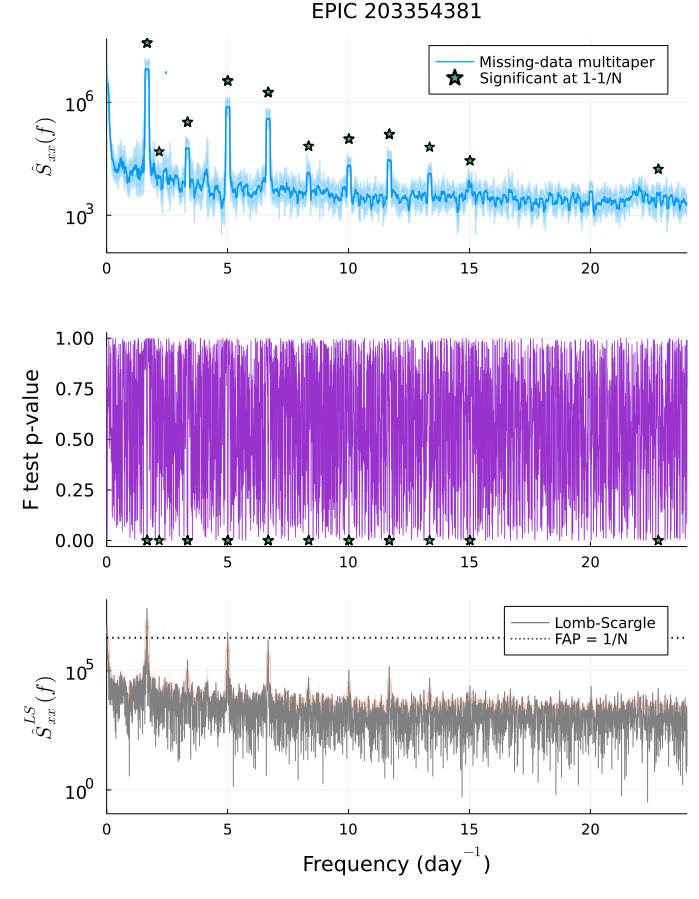}
    \caption{Power spectrum estimates from the time series $x_n$ plotted in Figure \ref{fig:scallop_data}. {Top:} The blue line shows the missing-data multitaper with $w^{(k)}_t$ from Figure \ref{fig:scallop_data}, with shaded 95\% confidence intervals. Green stars show signals with significance at least $1-1/N$ according to the F-test. {Middle:} The p-value associated with the F statistic at each frequency. Green stars have the same meaning as above. {Bottom:} The Lomb-Scargle periodogram plotted on top of the multitaper confidence intervals, with the dotted line showing the $1/N$ false alarm level.}
    \label{fig:scallop_spectrum}
\end{figure*}

{Figure \ref{fig:scallop_spectrum} shows the multitaper power spectrum estimate obtained from the entire dataset (top), the F-test p-value (middle), and the multitaper confidence intervals overplotted on the Lomb-Scargle periodogram with the $1/N$ white-noise false alarm threshold calculated using the \citet{baluev08} method (bottom). Green stars denote candidate signals for which the null hypothesis probability is $< 1/N$ according to the multitaper / F-test combination. The multitaper reveals a fundamental mode with $f_0 = 1.675$~day$^{-1}$, which is within 0.1\% of the rotation frequency reported by \citet{stauffer18}. The power spectrum estimate also contains a set of harmonics with $f = m f_0$, where $m = 2, 3, \ldots, 9$. Assuming the coronal slingshotting hypothesis is correct, the high-order harmonics reveal that the magnetic structure of the stellar corona is complex, and gas ejections have occurred at many longitudes. The magnetic field must be strong enough to retain some of the ejected gas inside a magnetosphere instead of allowing it to be expelled as a stellar wind \citep[e.g.][]{morin08}. There are also two oscillation candidates identified by the F-test that are not part of the rotation harmonic sequence. Their origins are unclear, though we roughly expect one false positive per power spectrum estimate with an F-test p-value cutoff of $1 / N$.}

{Figure \ref{fig:scallop_spectrum} highlights the value of a statistical test for harmonic components in the power spectrum that does not assume any particular noise model. Although spikes in the Lomb-Scargle periodogram are visible at the rotation frequency and harmonics $m = 2, 3, \ldots, 8$, only the signals at $m = 1$ and $m = 3$ rise above the $1/N$ false alarm level (though $m = 4$ is close). Multitaper allows the observers to meaningfully test the hypothesis that oscillatory components are present at high frequencies, even though the redness of the power spectrum ensures that only low-frequency signals can achieve statistical significance according to the white noise assumption. As an aside, although red noise is not obvious on a periodogram plot with a linear y-axis \citep[e.g.][]{murgas23, stalport23}, it is easily revealed as a negative slope on a semilog-y plot with frequency on the x-axis (see Appendix A of \citet{SDR2022} for a more detailed explanation of why power spectrum plots should have logarithmic y-axes).}

\begin{figure*}
    \centering
    \includegraphics[width=0.7\textwidth]{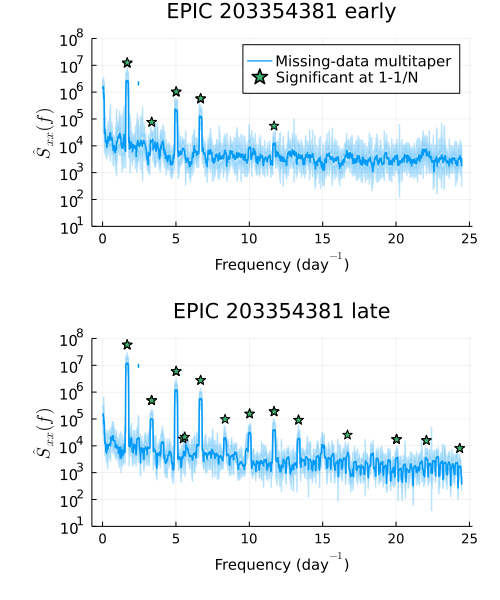}
    \caption{Multitaper power spectrum estimates from the early (top) and late (bottom) parts of the EPIC~203354381 K2 light curve. Green stars mark signals with F-test p-value $< 1/N$.}
    \label{fig:scallop_early_late_spectra}
\end{figure*}

{Figure \ref{fig:scallop_early_late_spectra} shows multitaper power spectrum estimates from the early (top) and late (bottom) parts of the dataset. While the fundamental rotation frequency plus harmonics $m = 2, 3, 4, 7$ are detected in the early data, the entire set of harmonics present in Figure \ref{fig:scallop_spectrum} does not appear. On the other hand, the later observations taken after $\operatorname{BJD} = 2456935$ contain harmonics $m = 2, 3, \ldots, 8, 10, 12$. All oscillations in the rotation and harmonic sequence have more power than in the early data. There are also candidate oscillations that are not associated with the rotation at $f = 5.48$~day$^{-1}$ and $f = 5.60$~day$^{-1}$, plus high-frequency signals with $f > 22$~day$^{-1}$. If these oscillations are real, their origin is unclear.}


{Figures \ref{fig:scallop_data}, \ref{fig:scallop_folded}, and \ref{fig:scallop_early_late_spectra} could indicate that at $\operatorname{BJD} = 2456935$, the stellar corona slingshotted a parcel of gas at a longitude that was formerly clear. The added complexity of the circumstellar environment manifested in the high-order harmonics present in the power spectrum of the late data.}


\section{Conclusions}
\label{sec:conclusions}

We have shown how to apply the missing-data multitaper power spectrum
estimator to spacecraft time series with regular observing cadence that have missing observations. The MDMT has the following advantages over the Lomb-Scargle periodogram:




\begin{itemize}

\item {Fewer false positives:} By averaging together $K$ independent power spectrum estimates, multitapering reduces the variance of the estimator by a factor of almost $1/K$. In addition, stronger serial correlations for classical and missing-data multitaper estimators produce fewer false positives than the (windowed) Schuster periodogram \citep{th14}. 
We find similar improvement over the generalized Lomb-Scargle periodogram in the science examples we show here (\S \ref{sec:kepler_diffrot} and \ref{sec:scallop}). 



\item {Bias suppression:} The {broadband} bias of a multitaper power spectrum estimate is limited to $(1 - \lambda^{(K-1)}) s_x^2$. We urge caution when interpreting Lomb-Scargle estimates: there is no guarantee that leakage from a large (true) oscillation is not causing spurious signals very far away from the main feature. We stress that this fact implies that \emph{when the data actually contain one or more large peaks, the number of spurious peaks may be more than is expected for white noise}, which can lead to false positives. {Furthermore, missing-data multitaper performs better than a periodogram at reproducing the shape and dynamic range of the power spectrum, which is crucial for studying broad-spectrum phenomena such as granulation and turbulence.}


\item {Minimal pseudowindowing:} Formally the MDMT spectral window is the same for all frequencies \citep{chave2019multitaper}, and in practice we have found that spectral window variation as a function of frequency is limited to $\sim 20$\%. Our results show that Lomb-Scargle pseudowindows can differ by an order of magnitude (\S \ref{subsec:windows}).

\item {True large peaks take on the rectangular shape of the spectral window,} which is given by a linear combination of tapers that are optimally large on the interval $(-\varpi,\varpi)$ and small everywhere else. The rectangular shape serves as a visual indicator of a purely sinusoidal oscillation in the power spectrum.

\item {F-testing:} To find statistically significant oscillations, an observer can use the F-test for harmonic components, which works directly on the complex eigencoefficients. The F-test can identify oscillations embedded in a red noise background that would not rise above a white noise-based false alarm threshold, such as the high-frequency modes in the EPIC 203354382 K2 dataset (\S \ref{sec:scallop}).

\item {Confidence intervals:} The precision of the MDMT can be quantified, which is important when the science goal requires estimating oscillation power (\S \ref{sec:kepler_diffrot}).

\end{itemize}


In summary, the missing-data multitaper power spectrum estimator is the best Fourier analysis tool for space missions such that return regularly spaced time series with gaps, such as {\it Kepler}, K2, and the upcoming Twinkle mission \citep{stotesbury22}.



 \section*{Acknowledgements}

The authors would like to thank Alan Chave for discussion on missing-data power spectrum estimators. This material is based upon work supported by the U.S. Department of Energy, 529 Office of Science, under contract number DE-AC02-06CH11357. Additional funding was provided by the Bartol Research Institute. 
{We thank the contribution of an anonymous reviewer for their remarks on an earlier version of this work.}

 The submitted manuscript has been created by UChicago Argonne, LLC, Operator of Argonne National Laboratory (“Argonne”). Argonne, a U.S. Department of Energy Office of Science laboratory, is operated under Contract No.\ DE-AC02-06CH11357. The U.S. Government retains for itself, and others acting on its behalf, a paid-up nonexclusive, irrevocable worldwide license in said article to reproduce, prepare derivative works, distribute copies to the public, and perform publicly and display publicly, by or on behalf of the Government. The Department of Energy will provide public access to these results of federally sponsored research in accordance with the DOE Public Access Plan. http://energy.gov/downloads/doe-public-access-plan

\section*{Software}
\label{sec:software}

All multitaper calculations in this article were performed with the \texttt{Julia} package Multitaper.jl \citep[][https://github.com/lootie/Multitaper.jl]{Haley2020}. The github repository contations tutorial \texttt{Jupyter} notebooks that demonstrate all multitaper functionality, the notebooks that perform the calculations in this article, and a quickstart guide to \texttt{Julia} for astronomers. For \texttt{python} users, the wrapper \texttt{multitaperpy} is available at https://github.com/lootie/multitaperpy/. This work also made use of LombScargle.jl \citep{lombscarglejl}, Astropy \citep{astropy:2013, astropy:2018}, and FINUFFT \citep{barnett20}.

\section*{Data Availability}

The K2 observations of EPIC~203354381 used in \S \ref{sec:scallop} can be accessed via \dataset[10.5281/zenodo.8436150]{https://doi.org/10.5281/zenodo.8436150}.

KIC 6102338 {\it Kepler} Q0, Q3, and Q16 PDCSAP flux data were accessed from the Mikulski Archive for Space Telescopes (MAST) at the Space Telescope Science Institute. The specific observations analyzed can be accessed via \dataset[doi:10.17909/0etb-4t73]{https://archive.stsci.edu/doi/resolve/resolve.html?doi=10.17909/0etb-4t73}.

CHEOPS timesteps from a 55~Cnc observing sequence were downloaded from the CHEOPS mission archive (https://cheops-archive.astro.unige.ch/archive\_browser/) and can be accessed via \dataset[10.5281/zenodo.8436180]{https://doi.org/10.5281/zenodo.8436180}. 


%

\bibliographystyle{mnras}
\bibliography{nfftbib,biblio}

\begin{thebibliography}{}
\makeatletter
\relax
\def\mn@urlcharsother{\let\do\@makeother \do\$\do\&\do\#\do\^\do\_\do\%\do\~}
\def\mn@doi{\begingroup\mn@urlcharsother \@ifnextchar [ {\mn@doi@} {\mn@doi@[]}}
\def\mn@doi@[#1]#2{\def\@tempa{#1}\ifx\@tempa\@empty \href {http://dx.doi.org/#2} {doi:#2}\else \href {http://dx.doi.org/#2} {#1}\fi \endgroup}
\def\mn@eprint#1#2{\mn@eprint@#1:#2::\@nil}
\def\mn@eprint@arXiv#1{\href {http://arxiv.org/abs/#1} {{\tt arXiv:#1}}}
\def\mn@eprint@dblp#1{\href {http://dblp.uni-trier.de/rec/bibtex/#1.xml} {dblp:#1}}
\def\mn@eprint@#1:#2:#3:#4\@nil{\def\@tempa {#1}\def\@tempb {#2}\def\@tempc {#3}\ifx \@tempc \@empty \let \@tempc \@tempb \let \@tempb \@tempa \fi \ifx \@tempb \@empty \def\@tempb {arXiv}\fi \@ifundefined {mn@eprint@\@tempb}{\@tempb:\@tempc}{\expandafter \expandafter \csname mn@eprint@\@tempb\endcsname \expandafter{\@tempc}}}

\bibitem[\protect\citeauthoryear{{Anglada-Escud{\'e}} et~al.,}{{Anglada-Escud{\'e}} et~al.}{2016}]{angladaescude16}
{Anglada-Escud{\'e}} G.,  et~al., 2016, \mn@doi [\nat] {10.1038/nature19106}, \href {https://ui.adsabs.harvard.edu/abs/2016Natur.536..437A} {536, 437}

\bibitem[\protect\citeauthoryear{{Astropy Collaboration} et~al.,}{{Astropy Collaboration} et~al.}{2013}]{astropy:2013}
{Astropy Collaboration} et~al., 2013, \mn@doi [\aap] {10.1051/0004-6361/201322068}, \href {https://ui.adsabs.harvard.edu/abs/2013A&A...558A..33A} {558, A33}

\bibitem[\protect\citeauthoryear{{Astropy Collaboration} et~al.,}{{Astropy Collaboration} et~al.}{2018}]{astropy:2018}
{Astropy Collaboration} et~al., 2018, \mn@doi [\aj] {10.3847/1538-3881/aabc4f}, \href {https://ui.adsabs.harvard.edu/abs/2018AJ....156..123A} {156, 123}

\bibitem[\protect\citeauthoryear{{Baliunas} et~al.,}{{Baliunas} et~al.}{1995}]{baliunas95}
{Baliunas} S.~L.,  et~al., 1995, \mn@doi [\apj] {10.1086/175072}, \href {https://ui.adsabs.harvard.edu/abs/1995ApJ...438..269B} {438, 269}

\bibitem[\protect\citeauthoryear{{Baluev}}{{Baluev}}{2008}]{baluev08}
{Baluev} R.~V.,  2008, \mn@doi [\mnras] {10.1111/j.1365-2966.2008.12689.x}, \href {https://ui.adsabs.harvard.edu/abs/2008MNRAS.385.1279B} {385, 1279}

\bibitem[\protect\citeauthoryear{{Barnes}}{{Barnes}}{2010}]{barnes10}
{Barnes} S.~A.,  2010, \mn@doi [\apj] {10.1088/0004-637X/722/1/222}, \href {https://ui.adsabs.harvard.edu/abs/2010ApJ...722..222B} {722, 222}

\bibitem[\protect\citeauthoryear{{Barnett}}{{Barnett}}{2020}]{barnett20}
{Barnett} A.~H.,  2020, arXiv e-prints, \href {https://ui.adsabs.harvard.edu/abs/2020arXiv200109405B} {p. arXiv:2001.09405}

\bibitem[\protect\citeauthoryear{Barnett, Magland  \& af Klinteberg}{Barnett et~al.}{2019}]{barnett19}
Barnett A.~H.,  Magland J.,   af Klinteberg L.,  2019, \mn@doi [SIAM Journal on Scientific Computing] {10.1137/18M120885X}, 41, C479

\bibitem[\protect\citeauthoryear{Bartlett}{Bartlett}{1948}]{bartlett48}
Bartlett M.~S.,  1948, Nature, 161

\bibitem[\protect\citeauthoryear{Blackman \& Tukey}{Blackman \& Tukey}{1958}]{blackman1958measurement}
Blackman R.~B.,  Tukey J.~W.,  1958, Bell System Technical Journal, 37, 185

\bibitem[\protect\citeauthoryear{Bloomfield}{Bloomfield}{2000}]{bloomfield00}
Bloomfield P.,  2000, Fourier analysis of time series: an introduction, 2nd ed..
Wiley series in probability and statistics. Applied probability and statistics section, Wiley, New York

\bibitem[\protect\citeauthoryear{{Boisse}, {Bonfils}  \& {Santos}}{{Boisse} et~al.}{2012}]{boisse2012}
{Boisse} I.,  {Bonfils} X.,   {Santos} N.~C.,  2012, \mn@doi [A\&A] {10.1051/0004-6361/201219115}, 545, A109

\bibitem[\protect\citeauthoryear{{Bourrier} et~al.,}{{Bourrier} et~al.}{2018}]{bourrier18}
{Bourrier} V.,  et~al., 2018, \mn@doi [\aap] {10.1051/0004-6361/201833154}, \href {https://ui.adsabs.harvard.edu/abs/2018A&A...619A...1B} {619, A1}

\bibitem[\protect\citeauthoryear{{Bragaglia}, {Greggio}, {Renzini}  \& {D'Odorico}}{{Bragaglia} et~al.}{1990}]{bragaglia90}
{Bragaglia} A.,  {Greggio} L.,  {Renzini} A.,   {D'Odorico} S.,  1990, \mn@doi [\apjl] {10.1086/185877}, \href {https://ui.adsabs.harvard.edu/abs/1990ApJ...365L..13B} {365, L13}

\bibitem[\protect\citeauthoryear{Brillinger}{Brillinger}{1975}]{brillinger75}
Brillinger D.~R.,  1975, Time Series, Data Analysis and Theory.
Holt, Rinehart, and Winston, New York

\bibitem[\protect\citeauthoryear{Bronez}{Bronez}{1988}]{bronez1988spectral}
Bronez T.~P.,  1988, IEEE Trans. on Acoustics, Speech, and Signal Processing, 36, 1862

\bibitem[\protect\citeauthoryear{Bronez}{Bronez}{1992}]{bronez92}
Bronez T.~P.,  1992, IEEE Trans. on Signal Processing, 40, 2941

\bibitem[\protect\citeauthoryear{Chave}{Chave}{2019}]{chave2019multitaper}
Chave A.~D.,  2019, Geophys. J. Inter., 218, 2165

\bibitem[\protect\citeauthoryear{{Chave}, {Luther}  \& {Thomson}}{{Chave} et~al.}{2019}]{chave2019ocean}
{Chave} A.~D.,  {Luther} D.~S.,   {Thomson} D.~J.,  2019, \mn@doi [Journal of Geophysical Research (Oceans)] {10.1029/2018JC014586}, \href {https://ui.adsabs.harvard.edu/abs/2019JGRC..124.2072C} {124, 2072}

\bibitem[\protect\citeauthoryear{{Chitta}, {van Ballegooijen}, {Rouppe van der Voort}, {DeLuca}  \& {Kariyappa}}{{Chitta} et~al.}{2012}]{chitta12}
{Chitta} L.~P.,  {van Ballegooijen} A.~A.,  {Rouppe van der Voort} L.,  {DeLuca} E.~E.,   {Kariyappa} R.,  2012, \mn@doi [\apj] {10.1088/0004-637X/752/1/48}, \href {https://ui.adsabs.harvard.edu/abs/2012ApJ...752...48C} {752, 48}

\bibitem[\protect\citeauthoryear{Cram{\'e}r}{Cram{\'e}r}{1940}]{cramer40}
Cram{\'e}r H.,  1940, Annals of Mathematics, 41, 215

\bibitem[\protect\citeauthoryear{Cressie}{Cressie}{1981}]{cressie81}
Cressie N.,  1981, J Roy Statist Soc, B, 43, 177

\bibitem[\protect\citeauthoryear{{Cumming}}{{Cumming}}{2004}]{cumming04}
{Cumming} A.,  2004, \mn@doi [\mnras] {10.1111/j.1365-2966.2004.08275.x}, \href {https://ui.adsabs.harvard.edu/abs/2004MNRAS.354.1165C} {354, 1165}

\bibitem[\protect\citeauthoryear{{Damasso} et~al.,}{{Damasso} et~al.}{2020}]{damasso20}
{Damasso} M.,  et~al., 2020, \mn@doi [Science Advances] {10.1126/sciadv.aax7467}, \href {https://ui.adsabs.harvard.edu/abs/2020SciA....6.7467D} {6, eaax7467}

\bibitem[\protect\citeauthoryear{{Dawson} \& {Fabrycky}}{{Dawson} \& {Fabrycky}}{2010}]{dawson10}
{Dawson} R.~I.,  {Fabrycky} D.~C.,  2010, \mn@doi [\apj] {10.1088/0004-637X/722/1/937}, \href {https://ui.adsabs.harvard.edu/abs/2010ApJ...722..937D} {722, 937}

\bibitem[\protect\citeauthoryear{{Delisle}, {Hara}  \& {S\'egransan}}{{Delisle} et~al.}{2020}]{delisle2020}
{Delisle} J.-B.,  {Hara} N.,   {S\'egransan} D.,  2020, \mn@doi [A\&A] {10.1051/0004-6361/201936906}, 638, A95

\bibitem[\protect\citeauthoryear{{Dodson-Robinson}, {Delgado}, {Harrell}  \& {Haley}}{{Dodson-Robinson} et~al.}{2022}]{SDR2022}
{Dodson-Robinson} S.~E.,  {Delgado} V.~R.,  {Harrell} J.,   {Haley} C.~L.,  2022, \mn@doi [\aj] {10.3847/1538-3881/ac52ed}, \href {https://ui.adsabs.harvard.edu/abs/2022AJ....163..169D} {163, 169}

\bibitem[\protect\citeauthoryear{Efron \& Stein}{Efron \& Stein}{1981}]{efronstein81}
Efron B.,  Stein C.,  1981, Ann. Statist., 9, 586

\bibitem[\protect\citeauthoryear{{Faria} et~al.,}{{Faria} et~al.}{2022}]{faria22}
{Faria} J.~P.,  et~al., 2022, \mn@doi [\aap] {10.1051/0004-6361/202142337}, \href {https://ui.adsabs.harvard.edu/abs/2022A&A...658A.115F} {658, A115}

\bibitem[\protect\citeauthoryear{{Fowler} et~al.,}{{Fowler} et~al.}{2010}]{fowler10}
{Fowler} J.~W.,  et~al., 2010, \mn@doi [\apj] {10.1088/0004-637X/722/2/1148}, \href {https://ui.adsabs.harvard.edu/abs/2010ApJ...722.1148F} {722, 1148}

\bibitem[\protect\citeauthoryear{{Frasca}, {Fr{\"o}hlich}, {Bonanno}, {Catanzaro}, {Biazzo}  \& {Molenda-{\.Z}akowicz}}{{Frasca} et~al.}{2011}]{frasca11}
{Frasca} A.,  {Fr{\"o}hlich} H.~E.,  {Bonanno} A.,  {Catanzaro} G.,  {Biazzo} K.,   {Molenda-{\.Z}akowicz} J.,  2011, \mn@doi [\aap] {10.1051/0004-6361/201116980}, \href {https://ui.adsabs.harvard.edu/abs/2011A&A...532A..81F} {532, A81}

\bibitem[\protect\citeauthoryear{{Fr{\"o}hlich}, {Frasca}, {Catanzaro}, {Bonanno}, {Corsaro}, {Molenda-{\.Z}akowicz}, {Klutsch}  \& {Montes}}{{Fr{\"o}hlich} et~al.}{2012}]{frohlich12}
{Fr{\"o}hlich} H.~E.,  {Frasca} A.,  {Catanzaro} G.,  {Bonanno} A.,  {Corsaro} E.,  {Molenda-{\.Z}akowicz} J.,  {Klutsch} A.,   {Montes} D.,  2012, \mn@doi [\aap] {10.1051/0004-6361/201219167}, \href {https://ui.adsabs.harvard.edu/abs/2012A&A...543A.146F} {543, A146}

\bibitem[\protect\citeauthoryear{{Gaia Collaboration}}{{Gaia Collaboration}}{2020}]{gaia3}
{Gaia Collaboration} 2020, VizieR Online Data Catalog, \href {https://ui.adsabs.harvard.edu/abs/2020yCat.1350....0G} {p. I/350}

\bibitem[\protect\citeauthoryear{{Garc{\'\i}a} et~al.,}{{Garc{\'\i}a} et~al.}{2011}]{garcia11}
{Garc{\'\i}a} R.~A.,  et~al., 2011, \mn@doi [\mnras] {10.1111/j.1745-3933.2011.01042.x}, \href {https://ui.adsabs.harvard.edu/abs/2011MNRAS.414L...6G} {414, L6}

\bibitem[\protect\citeauthoryear{Giordano \& contributors}{Giordano \& contributors}{2017}]{lombscarglejl}
Giordano M.,  contributors 2017, {JuliaAstro/LombScargle.jl: Compute Lomb-Scargle periodogram, suitable for unevenly sampled data}, \url {https://github.com/JuliaAstro/LombScargle.jl}

\bibitem[\protect\citeauthoryear{Gr{\"u}nbaum}{Gr{\"u}nbaum}{1981}]{grunbaum1981eigenvectors}
Gr{\"u}nbaum F.~A.,  1981, SIAM Journal on Algebraic Discrete Methods, 2, 136

\bibitem[\protect\citeauthoryear{Haley \& Anitescu}{Haley \& Anitescu}{2017}]{haley17}
Haley C.~L.,  Anitescu M.,  2017, \mn@doi [IEEE Signal Processing Letters] {10.1109/LSP.2017.2719943}, 24, 1696

\bibitem[\protect\citeauthoryear{Haley \& Geoga}{Haley \& Geoga}{2020a}]{multitaperpkg}
Haley C.~L.,  Geoga C.~J.,  2020a, \texttt{Multitaper.jl}: a {J}ulia library for multitaper nonparametric spectrum analysis, \url{http://bitbucket.org/clhaley/Multitaper.jl}

\bibitem[\protect\citeauthoryear{Haley \& Geoga}{Haley \& Geoga}{2020b}]{Haley2020}
Haley C.~L.,  Geoga C.~J.,  2020b, \mn@doi [Journal of Open Source Software] {10.21105/joss.02463}, 5, 2463

\bibitem[\protect\citeauthoryear{Harris}{Harris}{1978}]{harris78}
Harris F.~J.,  1978, Proceedings of the IEEE, 66, 51

\bibitem[\protect\citeauthoryear{{Jenkins} et~al.,}{{Jenkins} et~al.}{2010}]{jenkins10}
{Jenkins} J.~M.,  et~al., 2010, \mn@doi [\apjl] {10.1088/2041-8205/713/2/L87}, \href {https://ui.adsabs.harvard.edu/abs/2010ApJ...713L..87J} {713, L87}

\bibitem[\protect\citeauthoryear{{Kallinger} et~al.,}{{Kallinger} et~al.}{2010}]{kallinger10}
{Kallinger} T.,  et~al., 2010, \mn@doi [\aap] {10.1051/0004-6361/201015263}, \href {https://ui.adsabs.harvard.edu/abs/2010A&A...522A...1K} {522, A1}

\bibitem[\protect\citeauthoryear{Keiner, Kunis  \& Potts}{Keiner et~al.}{2009}]{keiner2009using}
Keiner J.,  Kunis S.,   Potts D.,  2009, ACM Transactions on Mathematical Software (TOMS), 36, 1

\bibitem[\protect\citeauthoryear{Koopmans}{Koopmans}{1995}]{koopmans95}
Koopmans L.~H.,  1995, in Koopmans L.~H.,  ed., Probability and Mathematical Statistics, The Spectral Analysis of Time Series.
Academic Press, San Diego, pp 294--353, \mn@doi{https://doi.org/10.1016/B978-012419251-5/50011-9}, \url {https://www.sciencedirect.com/science/article/pii/B9780124192515500119}

\bibitem[\protect\citeauthoryear{Lees}{Lees}{1995}]{lees95}
Lees J.~M.,  1995, \mn@doi [Geophysical Research Letters] {https://doi.org/10.1029/94GL03221}, 22, 513

\bibitem[\protect\citeauthoryear{Lomb}{Lomb}{1976}]{lomb76}
Lomb N.~R.,  1976, \apjs, 39, 447

\bibitem[\protect\citeauthoryear{{Luger}, {Agol}, {Kruse}, {Barnes}, {Becker}, {Foreman-Mackey}  \& {Deming}}{{Luger} et~al.}{2016}]{luger16}
{Luger} R.,  {Agol} E.,  {Kruse} E.,  {Barnes} R.,  {Becker} A.,  {Foreman-Mackey} D.,   {Deming} D.,  2016, \mn@doi [\aj] {10.3847/0004-6256/152/4/100}, \href {https://ui.adsabs.harvard.edu/abs/2016AJ....152..100L} {152, 100}

\bibitem[\protect\citeauthoryear{{Mamajek} \& {Hillenbrand}}{{Mamajek} \& {Hillenbrand}}{2008}]{mamajek08}
{Mamajek} E.~E.,  {Hillenbrand} L.~A.,  2008, \mn@doi [\apj] {10.1086/591785}, \href {https://ui.adsabs.harvard.edu/abs/2008ApJ...687.1264M} {687, 1264}

\bibitem[\protect\citeauthoryear{{Marsh}, {Dhillon}  \& {Duck}}{{Marsh} et~al.}{1995}]{marsh95}
{Marsh} T.~R.,  {Dhillon} V.~S.,   {Duck} S.~R.,  1995, \mn@doi [\mnras] {10.1093/mnras/275.3.828}, \href {https://ui.adsabs.harvard.edu/abs/1995MNRAS.275..828M} {275, 828}

\bibitem[\protect\citeauthoryear{Miller}{Miller}{1974}]{Miller74}
Miller R.~G.,  1974, Biometrika, 61, 1

\bibitem[\protect\citeauthoryear{{Morin} et~al.,}{{Morin} et~al.}{2008}]{morin08}
{Morin} J.,  et~al., 2008, \mn@doi [\mnras] {10.1111/j.1365-2966.2008.13809.x}, \href {https://ui.adsabs.harvard.edu/abs/2008MNRAS.390..567M} {390, 567}

\bibitem[\protect\citeauthoryear{{Mortier}, {Faria}, {Correia}, {Santerne}  \& {Santos}}{{Mortier} et~al.}{2015}]{mortier15}
{Mortier} A.,  {Faria} J.~P.,  {Correia} C.~M.,  {Santerne} A.,   {Santos} N.~C.,  2015, \mn@doi [\aap] {10.1051/0004-6361/201424908}, \href {https://ui.adsabs.harvard.edu/abs/2015A&A...573A.101M} {573, A101}

\bibitem[\protect\citeauthoryear{Mudelsee}{Mudelsee}{2014}]{mudelsee14}
Mudelsee M.,  2014, Spectral Analysis.
Springer International Publishing, Cham, pp 169--215, \mn@doi{10.1007/978-3-319-04450-7_5}, \url {https://doi.org/10.1007/978-3-319-04450-7_5}

\bibitem[\protect\citeauthoryear{Munk \& Hasselman}{Munk \& Hasselman}{1964}]{munk64}
Munk W.,  Hasselman K.,  1964, Super-resolution of Tides.
reprinted 1965 by University of Washington Press, pp 339--344

\bibitem[\protect\citeauthoryear{{Murgas} et~al.,}{{Murgas} et~al.}{2023}]{murgas23}
{Murgas} F.,  et~al., 2023, \mn@doi [\aap] {10.1051/0004-6361/202346692}, \href {https://ui.adsabs.harvard.edu/abs/2023A&A...677A.182M} {677, A182}

\bibitem[\protect\citeauthoryear{Ojeda \& Whitman}{Ojeda \& Whitman}{2002}]{ojeda02}
Ojeda G.~Y.,  Whitman D.,  2002, \mn@doi [Journal of Geophysical Research: Solid Earth] {https://doi.org/10.1029/2000JB000114}, 107, ETG 3

\bibitem[\protect\citeauthoryear{{Park}, {Lindberg}  \& {Vernon}}{{Park} et~al.}{1987}]{park87}
{Park} J.,  {Lindberg} C.~R.,   {Vernon} Frank~L. I.,  1987, \mn@doi [\jgr] {10.1029/JB092iB12p12675}, \href {https://ui.adsabs.harvard.edu/abs/1987JGR....9212675P} {92, 12,675}

\bibitem[\protect\citeauthoryear{{P{\'e}Rez-Gussiny{\'e}}, {Lowry}  \& {Watts}}{{P{\'e}Rez-Gussiny{\'e}} et~al.}{2007}]{perez07}
{P{\'e}Rez-Gussiny{\'e}} M.,  {Lowry} A.~R.,   {Watts} A.~B.,  2007, \mn@doi [Geochemistry, Geophysics, Geosystems] {10.1029/2006GC001511}, \href {https://ui.adsabs.harvard.edu/abs/2007GGG.....8.5009P} {8, Q05009}

\bibitem[\protect\citeauthoryear{Percival}{Percival}{1994}]{percival94}
Percival D.~B.,  1994, in Stanford J.~L.,  Vardeman S.~B.,  eds, Methods in Experimental Physics, Vol.~28, Statistical Methods for Physical Science.
Academic Press, pp 313--348, \mn@doi{https://doi.org/10.1016/S0076-695X(08)60261-6}, \url {https://www.sciencedirect.com/science/article/pii/S0076695X08602616}

\bibitem[\protect\citeauthoryear{Percival \& Walden}{Percival \& Walden}{2020}]{pw2020}
Percival D.~B.,  Walden A.~T.,  2020, Spectral analysis for univariate time series.
Cambridge {U}niversity {P}ress

\bibitem[\protect\citeauthoryear{{Plachy} et~al.,}{{Plachy} et~al.}{2021}]{plachy21}
{Plachy} E.,  et~al., 2021, \mn@doi [\apjs] {10.3847/1538-4365/abd4e3}, \href {https://ui.adsabs.harvard.edu/abs/2021ApJS..253...11P} {253, 11}

\bibitem[\protect\citeauthoryear{Prieto}{Prieto}{2022}]{prieto22}
Prieto G.~A.,  2022, \mn@doi [Seismological Research Letters] {10.1785/0220210332}, 93, 1922

\bibitem[\protect\citeauthoryear{Prieto, Parker, Thomson, Vernon  \& Graham}{Prieto et~al.}{2007}]{prieto07}
Prieto G.~A.,  Parker R.~L.,  Thomson D.~J.,  Vernon F.~L.,   Graham R.~L.,  2007, \mn@doi [Geophysical Journal International] {10.1111/j.1365-246X.2007.03592.x}, 171, 1269

\bibitem[\protect\citeauthoryear{Rayleigh}{Rayleigh}{1903}]{rayleigh1903}
Rayleigh L.,  1903, Philosophical Magazine, 41, 238

\bibitem[\protect\citeauthoryear{{Reinhold} \& {Arlt}}{{Reinhold} \& {Arlt}}{2015}]{reinhold15}
{Reinhold} T.,  {Arlt} R.,  2015, \mn@doi [\aap] {10.1051/0004-6361/201425337}, \href {https://ui.adsabs.harvard.edu/abs/2015A&A...576A..15R} {576, A15}

\bibitem[\protect\citeauthoryear{{Reinhold}, {Reiners}  \& {Basri}}{{Reinhold} et~al.}{2013}]{reinhold13}
{Reinhold} T.,  {Reiners} A.,   {Basri} G.,  2013, \mn@doi [\aap] {10.1051/0004-6361/201321970}, \href {https://ui.adsabs.harvard.edu/abs/2013A&A...560A...4R} {560, A4}

\bibitem[\protect\citeauthoryear{{Robertson} \& {Mechoso}}{{Robertson} \& {Mechoso}}{1998}]{robertson98}
{Robertson} A.~W.,  {Mechoso} C.~R.,  1998, \mn@doi [Journal of Climate] {10.1175/1520-0442(1998)011<2570:IADCIR>2.0.CO;2}, \href {https://ui.adsabs.harvard.edu/abs/1998JCli...11.2570R} {11, 2570}

\bibitem[\protect\citeauthoryear{{Saunders}, {Luger}  \& {Barnes}}{{Saunders} et~al.}{2019}]{saunders19}
{Saunders} N.,  {Luger} R.,   {Barnes} R.,  2019, \mn@doi [\aj] {10.3847/1538-3881/ab12e4}, \href {https://ui.adsabs.harvard.edu/abs/2019AJ....157..197S} {157, 197}

\bibitem[\protect\citeauthoryear{Scargle}{Scargle}{1982}]{scargle1982studies}
Scargle J.~D.,  1982, The Astrophysical Journal, 263, 835

\bibitem[\protect\citeauthoryear{{Schuster}}{{Schuster}}{1898}]{schuster1898}
{Schuster} A.,  1898, \mn@doi [Terrestrial Magnetism (Journal of Geophysical Research)] {10.1029/TM003i001p00013}, \href {https://ui.adsabs.harvard.edu/abs/1898TeMag...3...13S} {3, 13}

\bibitem[\protect\citeauthoryear{{Schuster}}{{Schuster}}{1906}]{schuster06}
{Schuster} A.,  1906, \mn@doi [Philosophical Transactions of the Royal Society of London Series A] {10.1098/rsta.1906.0016}, \href {https://ui.adsabs.harvard.edu/abs/1906RSPTA.206...69S} {206, 69}

\bibitem[\protect\citeauthoryear{Slepian}{Slepian}{1978}]{s78}
Slepian D.,  1978, Bell System Tech. J., 57, 1371

\bibitem[\protect\citeauthoryear{{Smith} et~al.,}{{Smith} et~al.}{2012}]{smith12}
{Smith} J.~C.,  et~al., 2012, \mn@doi [\pasp] {10.1086/667697}, \href {https://ui.adsabs.harvard.edu/abs/2012PASP..124.1000S} {124, 1000}

\bibitem[\protect\citeauthoryear{{Springford}, {Eadie}  \& {Thomson}}{{Springford} et~al.}{2020}]{springford20}
{Springford} A.,  {Eadie} G.~M.,   {Thomson} D.~J.,  2020, \mn@doi [Astronomical J.] {10.3847/1538-3881/ab7fa1}, \href {https://ui.adsabs.harvard.edu/abs/2020AJ....159..205S} {159, 205}

\bibitem[\protect\citeauthoryear{{Stalport} et~al.,}{{Stalport} et~al.}{2023}]{stalport23}
{Stalport} M.,  et~al., 2023, \mn@doi [arXiv e-prints] {10.48550/arXiv.2308.05669}, \href {https://ui.adsabs.harvard.edu/abs/2023arXiv230805669S} {p. arXiv:2308.05669}

\bibitem[\protect\citeauthoryear{{Stauffer} et~al.,}{{Stauffer} et~al.}{2017}]{stauffer17}
{Stauffer} J.,  et~al., 2017, \mn@doi [\aj] {10.3847/1538-3881/aa5eb9}, \href {https://ui.adsabs.harvard.edu/abs/2017AJ....153..152S} {153, 152}

\bibitem[\protect\citeauthoryear{{Stauffer} et~al.,}{{Stauffer} et~al.}{2018}]{stauffer18}
{Stauffer} J.,  et~al., 2018, \mn@doi [\aj] {10.3847/1538-3881/aaa19d}, \href {https://ui.adsabs.harvard.edu/abs/2018AJ....155...63S} {155, 63}

\bibitem[\protect\citeauthoryear{Stoica \& Sundin}{Stoica \& Sundin}{1999}]{stoicasundin99}
Stoica P.,  Sundin T.,  1999, Circuits Systems Signal Process., 18, 169

\bibitem[\protect\citeauthoryear{{Stotesbury} et~al.,}{{Stotesbury} et~al.}{2022}]{stotesbury22}
{Stotesbury} I.,  et~al., 2022, in {Coyle} L.~E.,  {Matsuura} S.,   {Perrin} M.~D.,  eds,  Society of Photo-Optical Instrumentation Engineers (SPIE) Conference Series Vol. 12180, Space Telescopes and Instrumentation 2022: Optical, Infrared, and Millimeter Wave. p. 1218033 (\mn@eprint {arXiv} {2209.03337}), \mn@doi{10.1117/12.2641373}

\bibitem[\protect\citeauthoryear{{Stumpe} et~al.,}{{Stumpe} et~al.}{2012}]{stumpe12}
{Stumpe} M.~C.,  et~al., 2012, \mn@doi [\pasp] {10.1086/667698}, \href {https://ui.adsabs.harvard.edu/abs/2012PASP..124..985S} {124, 985}

\bibitem[\protect\citeauthoryear{Swain \& Kirby}{Swain \& Kirby}{2003}]{swain03}
Swain C.~J.,  Kirby J.~F.,  2003, \mn@doi [Geophysical Research Letters] {https://doi.org/10.1029/2003GL017070}, 30

\bibitem[\protect\citeauthoryear{Thomson}{Thomson}{1982}]{thomson1982spectrum}
Thomson D.~J.,  1982, Proceedings of the IEEE, 70, 1055

\bibitem[\protect\citeauthoryear{Thomson}{Thomson}{1990a}]{thomson90}
Thomson D.,  1990a, \mn@doi [phtrslb] {10.1098/rsta.1990.0041}, 330, 601

\bibitem[\protect\citeauthoryear{Thomson}{Thomson}{1990b}]{t90B}
Thomson D.~J.,  1990b, Phil. Trans. A, 332, 539

\bibitem[\protect\citeauthoryear{Thomson}{Thomson}{1994}]{T94jk}
Thomson D.~J.,  1994, in Proceedings of ICASSP, VI, 73

\bibitem[\protect\citeauthoryear{Thomson \& Chave}{Thomson \& Chave}{1991a}]{thomson91}
Thomson D.,  Chave A.,  1991a, Advances in Spectrum analysis and Array processing, 1, 55

\bibitem[\protect\citeauthoryear{Thomson \& Chave}{Thomson \& Chave}{1991b}]{TC91}
Thomson D.~J.,  Chave A.~D.,  1991b, in Haykin S.,  ed., , Vol.~1, Advances in Spectrum Analysis and Array Processing.
Prentice-Hall, Upper Saddle River, NJ, Chapt.~2, pp 58--113

\bibitem[\protect\citeauthoryear{Thomson \& Haley}{Thomson \& Haley}{2014}]{th14}
Thomson D.~J.,  Haley C.~L.,  2014, \mn@doi [Proceedings of the Royal Society A: Mathematical, Physical and Engineering Sciences] {10.1098/rspa.2014.0101}, 470, 20140101

\bibitem[\protect\citeauthoryear{{Van Kooten} \& {Cranmer}}{{Van Kooten} \& {Cranmer}}{2017}]{vankooten17}
{Van Kooten} S.~J.,  {Cranmer} S.~R.,  2017, \mn@doi [\apj] {10.3847/1538-4357/aa93e2}, \href {https://ui.adsabs.harvard.edu/abs/2017ApJ...850...64V} {850, 64}

\bibitem[\protect\citeauthoryear{{VanderPlas}}{{VanderPlas}}{2018}]{vanderplas18}
{VanderPlas} J.~T.,  2018, \mn@doi [\apjs] {10.3847/1538-4365/aab766}, \href {https://ui.adsabs.harvard.edu/abs/2018ApJS..236...16V} {236, 16}

\bibitem[\protect\citeauthoryear{{Vanderspek}, {Doty}, {Fausnaugh}, {Villase\~{n}or}, {Jenkins}, {Berta-Thompson}, {Burke}  \& {Ricker}}{{Vanderspek} et~al.}{2018}]{tesshandbook}
{Vanderspek} R.,  {Doty} J.,  {Fausnaugh} M.,  {Villase\~{n}or} J.,  {Jenkins} J.,  {Berta-Thompson} Z.,  {Burke} C.,   {Ricker} G.,  2018, TESS Instrument Handbook, \url {https://archive.stsci.edu/missions/tess/doc/TESS_Instrument_Handbook_v0.1.pdf}

\bibitem[\protect\citeauthoryear{Vautard, Yiou  \& Ghil}{Vautard et~al.}{1992}]{vautard92}
Vautard R.,  Yiou P.,   Ghil M.,  1992, \mn@doi [Physica D: Nonlinear Phenomena] {https://doi.org/10.1016/0167-2789(92)90103-T}, 58, 95

\bibitem[\protect\citeauthoryear{Welch}{Welch}{1967}]{welch67}
Welch P.,  1967, IEEE Transactions on Audio and Electroacoustics, AU-15, 70

\bibitem[\protect\citeauthoryear{Yaglom}{Yaglom}{1962}]{yaglom62}
Yaglom A.~M.,  1962, An Introduction to the Theory of Stationary Random Functions.
Prentice Hall

\bibitem[\protect\citeauthoryear{{Zechmeister} \& {K{\"u}rster}}{{Zechmeister} \& {K{\"u}rster}}{2009}]{zechmeister09}
{Zechmeister} M.,  {K{\"u}rster} M.,  2009, \mn@doi [\aap] {10.1051/0004-6361:200811296}, \href {https://ui.adsabs.harvard.edu/abs/2009A&A...496..577Z} {496, 577}

\bibitem[\protect\citeauthoryear{{Zhou}, {Wang}, {Zhou}, {Li}  \& {Chan}}{{Zhou} et~al.}{2007}]{zhou07}
{Zhou} W.,  {Wang} X.,  {Zhou} T.~J.,  {Li} C.,   {Chan} J.~C.~L.,  2007, \mn@doi [Meteorology and Atmospheric Physics] {10.1007/s00703-007-0263-6}, \href {https://ui.adsabs.harvard.edu/abs/2007MAP....98..283Z} {98, 283}

\makeatother
\end{thebibliography}


\end{document}